\documentclass[12pt,onecolumn,draftcls]{IEEEtran}

\usepackage{cite}      
\usepackage[dvips]{graphicx}  
\usepackage{amssymb, amsmath}   
\usepackage{subfigure}
\usepackage{multirow}
\IEEEoverridecommandlockouts

\begin{document}
%
%
\newcommand{\sinc}[1]{\mbox{sinc}\left({#1}\right)}
\newcommand{\erf}[1]{\mbox{erf}\left({#1}\right)}
\newcommand{\erfc}[1]{\mbox{erfc}\left({#1}\right)}
\newcommand{\corr}[1]{\mbox{corr}\left({#1}\right)}
\newcommand{\cov}[1]{\mbox{cov}\left({#1}\right)}
\newcommand{\var}[1]{\mbox{var}\left({#1}\right)}
\newcommand{\sgn}[1]{\mbox{sgn}\left({#1}\right)}
\newcommand{\mat}[1]{\boldsymbol{#1}}


\title{On Reduced Complexity Soft-Output MIMO ML detection}

\author{Massimiliano~Siti, ~\IEEEmembership{Member,~IEEE,} and~Michael~P.~Fitz,
~\IEEEmembership{Senior~Member,~IEEE}%
\thanks{Massimiliano Siti is with the Advanced System Technologies group of STMicroelectronics. Michael P. Fitz is with the EE Department of the University of California Los Angeles. The work of Massimiliano Siti was supported by a
Marie Curie International Fellowship within the $6^{th}$ European
Community Framework Programme.}%
}

\maketitle

\begin{abstract}
In multiple-input multiple-output (MIMO) fading channels maximum
likelihood (ML) detection is desirable to achieve high
performance, but its complexity grows exponentially with the
spectral efficiency. The current state of the art in MIMO
detection is list decoding and lattice decoding. This paper
proposes a new class of lattice detectors that combines some of
the principles of both list and lattice decoding, thus resulting
in an efficient parallelizable implementation and near optimal
soft-ouput ML performance. The novel detector is called layered
orthogonal lattice detector (LORD), because it adopts a new
lattice formulation and relies on a channel orthogonalization
process. It should be noted that the algorithm achieves optimal
hard-output ML performance in case of two transmit antennas. For
two transmit antennas max-log bit soft-output information can be
generated and for greater than two antennas approximate max-log
detection is achieved. Simulation results show that LORD, in MIMO
system employing orthogonal frequency division multiplexing (OFDM)
and bit interleaved coded modulation (BICM) is able to achieve
very high signal-to-noise ratio (SNR) gains compared to practical
soft-output detectors such as minimum-mean square error (MMSE), in
either linear or nonlinear iterative scheme. Besides, the
performance comparison with hard-output decoded algebraic space
time codes shows the fundamental importance of soft-output
generation capability for practical wireless applications.
\end{abstract}


\section{Introduction} \label{sec_intro}
Wireless transmission through multiple antennas, also referred to
as multiple-input multiple-output (MIMO) radio, is currently
enjoying a great popularity as it is considered the technology
able to satisfy the ever increasing demand of high data rate
communications. In MIMO fading channels and in presence of
additive white Gaussian noise (AWGN), maximum-likelihood (ML)
detection is optimal \cite{VE:76}. A straightforward
implementation of the ML detector would require, for an uncoded
complex constellation of size {\it S} and $L_t$ transmit antennas,
an exhaustive search over all possible $S^{L_t}$ transmit
sequences, thus being prohibitively complex for high spectral
efficiencies. This observation justifies the intense interest in
reduced complexity, sub-optimal linear detectors like zero-forcing
(ZF) or minimum mean square error (MMSE) \cite{Verdu}. These
algorithms currently represent the practical state-of-the-art for
MIMO coded systems, as they can easily generate bit soft output
information for use with powerful coded modulations. It should be
noted that ZF and MMSE with spatial multiplexing offer diversity
of $L_r-L_t+1$, where $L_r$ is the number of receive antennas,
while optimum detection provides $L_r$ \cite{AZ:00}; thus, linear
detectors are highly suboptimal. Nonlinear detectors based on the
combination of linear detectors and spatially ordered
decision-feedback equalization (O-DFE) were proposed for V-BLAST
in \cite{WW:98},\cite{JF:99}; they offer some performance
improvement, but suffer noise enhancements due to nulling and
error propagation due to interference cancellation (IC). More
interesting for bit interleaved coded modulation (BICM) systems
are soft-output iterative MMSE and error correction decoding
strategies, in either "hard" IC (HIC) \cite{SM:03} and soft IC
(SIC) \cite{SH:02}, \cite{ZWDV:03} schemes. However they suffer
from latency and complexity disadvantages.

To our knowledge, the class of ML approaching algorithms is quite
limited. Two important families are the \emph{list-based
detectors}
\cite{CNG:00},\cite{YL-ZQL:02},\cite{WB:04},\cite{WB:05},\cite{SLK:03},
based on the combination of ML and DFE principles, and the
\emph{lattice decoders}, among those the {\it sphere decoder} (SD)
\cite{EV:99} is most well known.

The common idea of the list-based detectors (LD) is to divide the
streams to be detected into two groups: first, one or more
reference transmit streams are selected and a corresponding list
of candidate constellation symbols is determined; then, for each
sequence in the list, interference is cancelled from the received
signal and the remaining symbol estimates are determined by as
many sub-detectors operating on reduced size sub-channels. The IC
process is analogous to V-BLAST spatial DFE; the differences lie
in the criterion adopted to select the reference layer(s) and to
order the remaining ones, and in the fact that its initial symbol
estimate is replaced by a \textit{list} of candidates. If the list
includes all possible constellation points
\cite{CNG:00},\cite{YL-ZQL:02}, an initial stage of interference
nulling is not required; this interference nulling is still
performed if a reduced size sorted list is generated starting from
the ZF estimates \cite{WB:04}, \cite{WB:05}. The final
hard-decision sequence is selected by minimizing the Euclidean
distance (ED) metrics over the considered sequences. A
particularly interesting result was obtained searching all
possible {\it S} cases for a reference stream, or {\it layer}, and
adopting O-DFE for the remaining $L_t-1$ sub-detectors. If a
properly optimized layer ordering technique is utilized, numerical
results reported in \cite{YL-ZQL:02},\cite{WB:04} demonstrate that
the LD detector is able to achieve full receive diversity and a
degradation from ML performance of fractions of a dB. A notable
property is that this can be accomplished through a parallel
implementation, as the sub-detectors can operate independently.
The drawback is that the computational complexity is high as $L_t$
O-DFE detectors for $L_t-1$ sub-streams have to be computed. If
efficiently implemented \cite{WB:05}, it involves $O(L_t^4)$
complexity. Another major shortcoming of the prior work in list
based detection is the absence of an algorithm to produce soft bit
metrics for use in coding and decoding algorithms.

Lattice detectors use the linear nature of the MIMO channel to form a
reduced complexity ML search. Lattice detectors are suitable for systems
whose input-output relation can be represented as a real-domain linear
model
\begin{equation}
\textbf{y}_r=\textbf{Bx}_r+\textbf{n}_r=\textbf{s}_r+\textbf{n}_r
\label{real_sys_eq}
\end{equation}
where the information sequence $\textbf{x}_r$ is uniformly
distributed over a discrete finite set $\mathcal{C} \subset R^m$,
$\textbf{n}_r\in R^n$ represents the noise vector, \textbf{B} is a
$n \times m$ real matrix where $n=2L_r$ and $m=2L_t$. The output
signal vector $\textbf{s}_r \in \Lambda\subset{R^{n}}$.
$\textbf{B}$ represents the channel mapping of the transmit
signals into the $m$-dimensional {\it lattice} $\Lambda$ and is
also referred to as the {\it lattice generator} matrix. If the
noise components are independent and identically distributed
(i.i.d.) zero-mean Gaussian random variables (RVs) with a common
variance, as typical of communication systems, then the ML
decoding rule corresponds to solving the minimization problem:
\begin{equation}
\hat{\textbf{x}}_r=\arg \min_{\textbf{x}_r\in \, \mathcal{C}}
\|\textbf{y}_r-\textbf{Bx}_r\|^2 \label{MLd}
\end{equation}
where $\|\cdot\|$ denotes the vector norm. The problem in (\ref{MLd})
is a constrained version of the {\it closest vector problem} (CVP)
in lattice theory. A survey of closest point search methods has
been presented in \cite{AEZ:02}. It should be noted that in the
general case the model of (\ref{real_sys_eq}) is still valid if a
general encoder matrix $\textbf{G}\in R^{\,m\times m}$ is
considered such that:
\begin{equation}
\textbf{x}_r=\textbf{Gu}_r \label{enc_sys}
\end{equation}
where $\textbf{u}_r \in \mathcal{U} \subset R^m$ is the
information symbol sequence and $\textbf{x}_r$ is the transmit
{\it codeword}. In this case the lattice generator matrix becomes
\textbf{BG}. In the rest of the paper we will refer to (\ref
{real_sys_eq}) with no loss of generality. The lattice formulation
for MIMO wireless systems was described in \cite{DCB:00} in case
of quadrature amplitude modulation (QAM) digitally modulated
transmitted symbols.

SD can attain ML performance with significantly reduced
complexity. Some efficient variants of the algorithm are
summarized in \cite{DGC:03}. SD presents a number of
disadvantages, which can be summarized as follows:
\begin{itemize}
\item[-] As the in-phase (I) and quadrature-phase (Q) components
of the digitally modulated QAM symbols are searched in a serial
fashion, SD is is not suitable for a parallel VLSI implementation.
As a support to our claim, some papers have described the SD
operations in terms of tree search and recently, the equivalence
between the SD and the sequential decoder has been established
rigorously \cite{MGDC:05}. \item[-] The number of lattice points
to be searched is a random variable, sensitive to the channel and
noise realizations, and to the initial radius. This implies a
non-deterministic complexity and latency, not desirable for
real-time high-data rate applications. Further, SD complexity has
often been referred to as polynomial but a recent work
\cite{JO:05} shows that SD remains an efficient solution only for
problems of moderate size, and for SNR values in given ranges.
This motivated the proposal of additional optional front-end
processing to expedite the lattice search. However,
\textit{lattice reduction} (LR) techniques such as the
Lenstra-Lenstra-Lov\'{a}sz \cite{LLL:82}, \cite{SE:94},
\cite{AEZ:02} did not prove very useful to solve the constrained
ML problem (\ref{MLd}), as reported in \cite{DGC:03}, because they
distort the original lattice and boundary control becomes
difficult. These lattice reduction techniques are useful in
conjunction with additional processing stage, i.e. the MMSE
``generalized decision feedback equalizer'' (MMSE-GDFE)
\cite{MGDC:05}. When channel is slowly varying, these front-end
processings are very effective on reducing the overall complexity.
However, if channel changes significantly from block to block,
front-end processings can have a significant impact on the overall
complexity. \item[-] Generation of soft output metrics is not easy
with known lattice decoding algorithms. A solution was proposed in
\cite{HTB:03} and successively refined in \cite{BGF:03}, where bit
log-likelihood ratios (LLR) are computed based on a "candidate
list" of sequences. No simple rule to determine the optimal size
of such list was proposed; simulation results show that it can be
very high (thousands of lattice points). This may nullify the
benefits of using SD, in terms of complexity, as also evidenced in
\cite{ZRF:04}. This observation is one of the main motivations of
this work. Also, in case of use of LR techniques the problem of
soft-output generation becomes prohibitively complex.
\end{itemize}

It should be noted that nulling and cancelling or equivalently
ZF-DFE, besides being the core of the O-DFE algorithm
\cite{JF:99}, is also an important part of the SD operations.
ZF-DFE can be efficiently implemented through a QR decomposition
(QRD) of the channel matrix, as shown in
\cite{SK:99},\cite{CNG:00} and \cite{DGC:03} for O-DFE and SD
respectively.

In an attempt to retain the advantages of LD and SD algorithms at
the same time addressing their main drawbacks, a novel MIMO
lattice detector is proposed in this paper.  This algorithm is
given the name layered orthogonal lattice detector (LORD)
\cite{MSF:05},\cite{MSF2:05}.

Similarly to SD, LORD consists of three different stages. First,
the system is represented through a proper lattice formulation but
different than the only one proposed for SD \cite{DCB:00}. Second
an efficient preprocessing of the channel matrix is implemented
for ZF-DFE. While a standard QRD could be employed without
altering the performance of the detection algorithm, a more
computationally efficient Gram-Schmidt orthogonalization (GSO)
process is outlined in this work. The last stage is the lattice
search, which involves finding a proper subset of transmit
sequences to solve the problem (\ref{MLd}). The number of lattice
points to be searched is linear in the number of transmit antennas
and is easily modified to provide soft output bit metrics. The
innovative concepts compared to SD, and already embedded in LD
\cite{YL-ZQL:02}, is that the search of the lattice points can be
accomplished in a parallel fashion, and their number is fully
deterministic. LORD achieves a huge complexity reduction over the
exhaustive-search ML algorithm and, as proven via numerical
results, also obtains a better complexity-performance tradeoff
than SD.

The proposed GSO technique avoids computing a complete QRD
multiple times if the channel columns are permuted, as clear from
the sequel. Compared to the best performing and efficiently
implemented LD ("B-Chase" detector \cite{WB:05}) that relies on
multiple QRDs, LORD algorithms has the following advantages. Its
preprocessing is less complex - $O(L_t^3)$ for $L_t=L_r$, instead
of $O(L_t^4)$); LORD generates reliable bit soft output
information; it does not require any particular ordering scheme
yet still retaining near-ML performance in BICM systems.
Nevertheless, it should be noted that concepts like reduced
complexity lattice search and optimal ordering schemes can be
applied to LORD as well, with proper adaptations to real-domain;
the explanation of these ideas is deferred to later works.

For two transmit antennas, LORD achieves ML hard-output
demodulation and is able to compute optimal (max-log) bit LLRs.
For more than two transmit antennas, the algorithm is suboptimal
but still near-ML in BICM systems and its gain over MMSE-based
linear and iterative nonlinear detectors actually increases with
the dimensionality of the problem, thanks to a good exploitation
of receive diversity. As shown later in this paper, even one
single stage of LORD processing performs better than several
iterations of MMSE-SIC in various orthogonal frequency division
multiplexing (OFDM) BICM systems, with clear latency advantages.
Iterative decoding and LORD detection schemes represent a
promising topic for future work. Overall, these results suggest
that soft-output MIMO near-ML detection, so far considered as
computationally intractable for real-time high-data rate
applications, can become a viable technique for next generation
wireless communication systems.

To conclude this section, it should be mentioned that no efficient
soft-output ML decoding strategy has been proposed so far for full
diversity full data rate algebraic space-time codes (STCs) like
the Golden Codes (GC) \cite{BRV:05}. The performance comparison of
layered BICM systems and uncoded GC provided in Section
\ref{sec_performance} shows that soft-output ML detection and ECCs
are essential in order to exploit the high-data rate and high link
robustness promised by MIMO for next generation wireless
applications (like wireless local area networks (WLANs),
undergoing standardization as IEEE 802.11n \cite{80211n:spec}).

The rest of the paper is organized as follows. In Section
\ref{sec_systemmodel} the system notation and the novel lattice
formulation are introduced. Section \ref{sec_LORD_2TX} is
concerned with the description of the stages of LORD for two
transmit antennas because LORD is optimal in this case. In Section
\ref{sec_preproc_2TX} an efficient preprocessing algorithm is
described; Section \ref{sec_latsearch_2TX} focuses on the lattice
search, and \ref{sec_LLR_2TX} deals with the bit soft output
generation. Section \ref{sec_LORD_NTX} and its subsections include
a formulation of LORD suitable for any number of transmit
antennas. Section \ref{sec_performance} confirms through numerical
results that LORD provides an excellent performance-complexity
tradeoff. Finally, some concluding remarks are reported in Section
\ref{sec_conclusion}.

\section{System Model and Lattice Representation} \label{sec_systemmodel}

We consider a MIMO communication system with $L_t$ transmit and
$L_r$ receive antennas, and a frequency nonselective fading channel.
We also assume the receiver has perfect knowledge of the channel
state and each receive antenna has a matched filter to the pulse
shape. Then the complex baseband received signal $\textbf{y}_c =
(Y_{c1} \, \ldots Y_{cL_r})^{T}$ is given by:
\begin{equation}
\textbf{y}_c=\sqrt{\frac{E_s}{L_t}} \,
\textbf{H}_c\textbf{x}_c+\textbf{n}_c \label{cmplx_sys_eq}
\end{equation}
where the input signal $\textbf{x}_c=(X_{c1} \, \ldots
X_{cL_t})^{T}$ is the QAM or phase shift keying (PSK)\footnote{It
should be noted that a single version of SD cannot handle both QAM
and PSK transmit symbols; a modified SD \cite{HTB:03} has been
proposed for the latter case. LORD can be easily adapted to handle
PSK modulations and requires only a minor modification in the
demodulation section.} complex information symbol vector, $E_s$ is
the energy per transmitted symbol (under the hypothesis that the
average constellation energy is $E[|X_{cj}|^2]=1$), \textbf{n}
$=(N_{c1} \, \ldots N_{cL_r})^{T}$ is the $L_r \times 1$ complex
white Gaussian noise (AWGN) sample vector, $\textbf{H}_c$ is the
$L_r \times L_t$ complex channel matrix. The entries of
$\textbf{H}_c$ are the i.i.d. complex path gains $H_{c\,ji}\sim
\mathcal{N}_c(0,1)$ from transmit antenna $i$ to receive antenna
$j$. At receive antenna $j$, the corrupting i.i.d. noise samples
are $N_{cj}\sim \mathcal{N}_c(0,N_0)$. As it will prove useful in
the following, the $i^{th}$ column of $\textbf{H}_c$ is denoted as
$\textbf{h}_{ci}$. Equation (\ref{cmplx_sys_eq}) is assumed to be
valid per each OFDM tone if a MIMO-OFDM system and frequency
selective channels are considered.

This paper assumes QAM modulation and derives a real lattice formulation.
As a variant to the traditional lattice formulation \cite{DCB:00}, the
system (\ref{cmplx_sys_eq}) can be translated into the form
(\ref{real_sys_eq}) performing appropriate scaling and ordering
the I and Q of the complex entries as follows:
\begin{eqnarray}
\textbf{x}_r &=& \left[X_{1,I}, \ X_{1,Q}, \ldots \, X_{L_t,I}, \
X_{L_t,Q}\right]^T \nonumber \\ &=& \left[x_1,
\ldots \, x_{2L_t}\right]^T\\
\textbf{y}_r &=& \left[Y_{1,I}, \ Y_{1,Q}, \ldots \, Y_{L_r,I}, \
Y_{L_r,Q}\right]^T\\ \textbf{n}_r &=& \left[N_{1,I}, \ N_{1,Q},
\ldots \,  N_{L_r,I}, \ N_{L_r,Q}\right]^T.
\end{eqnarray}
Then (\ref{cmplx_sys_eq}) can be re-written as:
\begin{equation}
\textbf{y}_r=\sqrt{\frac{E_{s}}{L_t}} \, \textbf{H}_r \textbf{x}_r
+ \textbf{n}_r = \sqrt{\frac{E_{s}}{L_t}} \,
\left[\begin{array}{ccc} \textbf{h}_1, \ldots , \,
\textbf{h}_{2L_t}
\end{array} \right] \, \textbf{x}_r + \textbf{n}_r. \label{lateqn}
\end{equation}
Each pair of columns ($\textbf{h}_{2k-1}, \ \textbf{h}_{2k}$),
$k=\{1,\ldots L_t\}$ of the real channel matrix $\textbf{H}_r$ has
the form:
\begin{eqnarray}
\textbf{h}_{2k-1} = \left[ \Re{\left[H_{1k}\right]}, \,
\Im{\left[H_{1k}\right]}, \ldots , \, \Re{\left[H_{L_rk}\right]},
\, \Im{\left[H_{L_rk}\right]} \right]^T\\
\textbf{h}_{2k} = \left[ -\Im{\left[H_{1k}\right]}, \,
\Re{\left[H_{1k}\right]}, \ldots , \, -\Im{\left[H_{L_rk}\right]},
\, \Re{\left[H_{L_rk}\right]} \right]^T
 \label{hcol}
\end{eqnarray}
As a direct consequence of this formulation, they are pairwise
orthogonal, i.e.
\begin{equation}
\textbf{h}_{2k-1}^T\textbf{h}_{2k}=0 \nonumber
\end{equation}
where $k=\{1,\ldots , L_t\}$. This property will prove to be
essential for LORD simplified demodulation. Other useful relations
are:
\begin{eqnarray}\label{h_norms}
\left\|\textbf{h}_{2k-1}\right\|^2&=&\left\|\textbf{h}_{2k}\right\|^2\\
\textbf{h}_{2k-1}^T\textbf{h}_{2j-1}&=&\textbf{h}_{2k}^T\textbf{h}_{2j},
\quad
\textbf{h}_{2k-1}^T\textbf{h}_{2j}=-\textbf{h}_{2k}^T\textbf{h}_{2j-1}
\nonumber
\end{eqnarray}
where $k,j=\{1,\ldots , L_t\}$ and $k\neq j$.

\section{LORD Algorithm - case of two transmit antennas} \label{sec_LORD_2TX}
This section is concerned with the derivation of LORD algorithm
for the case of $L_t=2$ transmit antennas. The two transmit
antenna case is called out separately because in this case LORD is
optimal. After the system is represented in the real-domain
through the novel I and Q ordering, the proposed lattice detection
algorithm requires two additional stages: preprocessing and
lattice search. The purpose of preprocessing is to turn the MIMO
channel into an upper triangular system. The proposed
transformation is a computationally efficient alternate to a QRD
as the normalizations are performed after the channel
orthogonalization is completed, although a standard QRD would not
impair the demodulation and performance properties of the
algorithm. 

\subsection{The preprocessing algorithm} \label{sec_preproc_2TX}
The channel matrix can be represented as
\begin{equation}
{\bf H}_r={\bf Q} {\bf R} {\bf \Lambda}_q
\end{equation}
for $L_r\ge 2$. To show this it is noted that an
$2L_t
\times 2L_r$ orthogonal matrix can be defined
\begin{equation}
{\bf Q}=\left[\begin{array}{ccc} \textbf{h}_1 \quad \textbf{h}_2
\quad \textbf{q}_3 \quad \textbf{q}_4 \end{array} \right]
\end{equation}
where
\begin{eqnarray}
\textbf{q}_3 &=&
\|\textbf{h}_1\|^2\textbf{h}_3-(\textbf{h}_1^T\textbf{h}_3)\textbf{h}_1-
(\textbf{h}_2^T\textbf{h}_3)\textbf{h}_2\\
\textbf{q}_{4} &=&
\|\textbf{h}_1\|^2\textbf{h}_4-(\textbf{h}_1^T\textbf{h}_4)\textbf{h}_1-
(\textbf{h}_2^T\textbf{h}_4)\textbf{h}_2.
\end{eqnarray}
Then, remembering (\ref{h_norms}) one has:
\begin{equation} {\bf Q}^T{\bf
Q}=\mbox{diag}\left[\left\|\textbf{h}_1\right\|^2,
\left\|\textbf{h}_1\right\|^2, \left\|\textbf{q}_3\right\|^2,
\left\|\textbf{q}_3\right\|^2 \right].
\end{equation}
It can also be written that
\begin{equation}
\left\|\textbf{q}_3\right\|^2=\left\|\textbf{h}_1\right\|^2\left(\left\|\textbf{h}_3\right\|^2
\left\|\textbf{h}_1\right\|^2
-\left(\textbf{h}_1^T\textbf{h}_3\right)^2-\left(\textbf{h}_2^T\textbf{h}_3\right)^2\right)=
\left\|\textbf{h}_1\right\|^2 r_3
\end{equation}
where by definition
\begin{equation}
\label{r3} r_3=\left\|\textbf{h}_3\right\|^2
\left\|\textbf{h}_1\right\|^2
-\left(\textbf{h}_1^T\textbf{h}_3\right)^2-\left(\textbf{h}_2^T\textbf{h}_3\right)^2.
\end{equation}
By defining the $2L_t \times 2L_t$ upper triangular matrix
\begin{equation}
{\bf R}=\left[\begin{array}{cccc}
1 & 0 & \textbf{h}_1^T\textbf{h}_3 & \textbf{h}_1^T\textbf{h}_4\\
0 & 1 & \textbf{h}_2^T\textbf{h}_3 & \textbf{h}_2^T\textbf{h}_4\\
0 & 0 & 1 & 0\\
0 & 0 & 0 & 1\end{array}\right]
\end{equation}
and the $2L_t \times 2L_t$ diagonal matrix
\begin{equation}
{\bf \Lambda}_q=\mbox{diag}\left[1, 1, \
\left\|\textbf{h}_1\right\|^{-2}, \left\|\textbf{h}_1\right\|^{-2}
\right]
\end{equation}
the original real channel matrix can be decomposed as
\begin{equation}
{\bf H}_r={\bf Q}{\bf R}{\bf \Lambda}_q.
\end{equation}
The linear preprocessing proposed in this paper is given as
\begin{equation}
\tilde{\textbf{y}}_r={\bf Q}^T \textbf{y}_r
\end{equation}
where all values of ${\bf Q}$ are simple functions of the known channel
coefficients. The signal model after preprocessing is given as
\begin{equation}
\tilde{\textbf{y}}_r=\sqrt{\frac{E_{s}}{2}}
\, {\bf \tilde{R}} \textbf{x}_r + {\bf Q}^T
\textbf{n}_r=\sqrt{\frac{E_{s}}{2}} \, \tilde{{\bf R}}
\textbf{x}_r + \tilde{\textbf{n}}_r \label{equivsys1}
\end{equation}
where
\begin{equation} \label{eq:rtilde}
\tilde{{\bf R}}={\bf Q}^T{\bf Q}{\bf R}{\bf
\Lambda}_q=\left[\begin{array}{cccc}
\left\|\textbf{h}_1\right\|^2 & 0 & \textbf{h}_1^T\textbf{h}_3 & \textbf{h}_1^T\textbf{h}_4\\
0 & \left\|\textbf{h}_1\right\|^2 & \textbf{h}_2^T\textbf{h}_3 & \textbf{h}_2^T\textbf{h}_4\\
0 & 0 & r_3 & 0\\
0 & 0 & 0 & r_3\end{array}\right].
\end{equation}
is in the desired upper triangular form for lattice demodulation
algorithms. System (\ref{equivsys1}) is the real-domain lattice
system equation the detection algorithm LORD uses, and is in the
form of (\ref{real_sys_eq}). The noise vector in the triangular
model still has independent components but the components have
unequal variances, i.e.,
\begin{equation}
{\bf
R}_{\tilde{n}_r}=E\left[\tilde{\textbf{n}}_r\tilde{\textbf{n}}_r^T\right]
= \frac{N_0}{2}\,\mbox{diag}\left[\left\|\textbf{h}_1\right\|^2, \
\left\|\textbf{h}_1\right\|^2, \ \left\|\textbf{h}_1\right\|^2
r_3, \ \left\|\textbf{h}_1\right\|^2 r_3 \right].
\end{equation}
The advantageous characteristic of the model formulation is that
$\tilde{R}_{12}=\tilde{R}_{34}=0$, i.e., each of the I and Q
components of each transmitted signal are broken into orthogonal
dimensions and can be searched in an independent fashion.

As a further observation, all parameters needed in this
triangularized model are a function of eight variables. Four of
the variables are functions of the channel only, i.e.,
\begin{equation}
\sigma^2_1=\left\|\textbf{h}_1\right\|^2 \qquad
\sigma^2_2=\left\|\textbf{h}_3\right\|^2 \qquad
s_1=\textbf{h}_1^T\textbf{h}_3 \qquad
s_2=\textbf{h}_1^T\textbf{h}_4.
\end{equation}
and four are functions of the channel and the observations, i.e.,
\begin{equation}
V_1=\textbf{h}_1^T\textbf{y}_r \qquad
V_2=\textbf{h}_2^T\textbf{y}_r \qquad
V_3=\textbf{h}_3^T\textbf{y}_r \qquad
V_4=\textbf{h}_4^T\textbf{y}_r.
\end{equation}
Also, equalities (\ref{h_norms}) imply that the $2 \times 2$
matrix in the upper right corner of ${\bf \tilde{R}}$ is a
rotation matrix. Specifically the required results for the upper
triangular formulation is
\begin{equation} \label{eq:utriangular}
\tilde{\textbf{y}}_r=\left[\begin{array}{c}
\tilde{y_1} \\
\tilde{y_2} \\
\tilde{y_3} \\
\tilde{y_4}\end{array} \right] = \left[\begin{array}{c}
V_1 \\
V_2 \\
\sigma^2_1V_3-s_1V_1+s_2V_2 \\
\sigma^2_1V_4-s_2V_1-s_1V_2\end{array}\right] \quad {\bf
\tilde{R}}=\left[\begin{array}{cccc}
\sigma^2_1 & 0 & s_1 & s_2\\
0 & \sigma^2_1 & -s_2 & s_1\\
0 & 0 & \sigma^2_1\sigma^2_2-s_1^2-s_2^2 & 0 \\
0 & 0 & 0 & \sigma^2_1\sigma^2_2-s_1^2-s_2^2
\end{array}\right].
\end{equation}
This formulation results in a preprocessing complexity (expressed
in terms of real multiplications, RMs) that is $O(16L_r+9)$.

As it will prove useful when dealing with soft output generation,
we also notice that shifting the ordering of the transmit antennas
results in a similar model:
\begin{equation}
\tilde{\textbf{y}}_s=\sqrt{\frac{E_{s}}{2}} \, \tilde{{\bf R}}_s
\textbf{x}_s + \tilde{\textbf{n}}_s \label{shift_equivsys}
\end{equation}
where
\begin{equation}
\tilde{\textbf{y}}_s=\left[\begin{array}{c}
\tilde{y_{s1}} \\
\tilde{y_{s2}} \\
\tilde{y_{s3}} \\
\tilde{y_{s4}}\end{array} \right]=\left[\begin{array}{c}
V_3 \\
V_4 \\
\sigma^2_2V_1-s_1V_3-s_2V_4 \\
\sigma^2_2V_2+s_2V_3-s_1V_4
\end{array}\right]
\, {\bf \tilde{R}}_s=\left[\begin{array}{cccc}
\sigma^2_2 & 0 & s_1 & -s_2\\
0 & \sigma^2_2 & s_2 & s_1\\
0 & 0 & \sigma^2_1\sigma^2_2-s_1^2-s_2^2 & 0 \\
0 & 0 & 0 & \sigma^2_1\sigma^2_2-s_1^2-s_2^2
\end{array}\right]
\label{shifteq}
\end{equation}
\begin{equation}
{\bf
R}_{\tilde{n}_s}=E\left[\tilde{\textbf{n}}_s\tilde{\textbf{n}}_s^T\right]
= \frac{N_0}{2}\,\mbox{diag}\left[\left\|\textbf{h}_3\right\|^2, \
\left\|\textbf{h}_3\right\|^2, \ \left\|\textbf{h}_3\right\|^2
r_3, \ \left\|\textbf{h}_3\right\|^2 r_3 \right]
\end{equation}
and $\textbf{x}_s=\left[x_3 \, x_4 \, x_1 \, x_2 \right]^T$.

Finally, we observe that there is an interesting relationship
between the triangularized model parameters and the complex
channel coefficients. First it should be noted that
$\sigma^2_1=\left|\textbf{h}_{c1}\right|^2$ and that
$\sigma^2_2=\left|\textbf{h}_{c2}\right|^2$. Secondly, the sample
cross correlation between the gains for transmit antenna 1 and
transmit antenna 2 is given as
\begin{equation}
\label{eq:chancorr1} \textbf{h}_{c2}^H\textbf{h}_{c1}=s_1+js_2.
\end{equation}
A sample crosscorrelation coefficient can be defined as
\begin{equation}\label{eq:chancorr2}
\rho_{12}=\frac{\textbf{h}_{c2}^H\textbf{h}_{c1}}{\sqrt{\sigma^2_1\sigma^2_2}}.
\end{equation}
Using (\ref{eq:chancorr2}), formula (\ref{r3}) can be written as
\begin{equation}
r_3=\sigma^2_1\sigma^2_2\left(1-\left|\rho_{12}\right|^2\right).
\end{equation}
It is apparent that when $L_r$ gets large the magnitude of
$\rho_{12}$ will go to zero and the MIMO detection problem for
each antenna will become completely decoupled.

\subsection{Lattice search and demodulation} \label{sec_latsearch_2TX}
The system equations defined in Section~\ref{sec_preproc_2TX} lead
naturally to a simplified yet optimal ML demodulation. Consider a
PSK or QAM constellation of size {\it S}. The discussion in this
paper will assume that $(M^2)$-QAM modulation is used on each
antenna but a generalization to any linear modulation is possible.
The optimum ML word demodulator (\ref{MLd}) would have to compute
the ML metric for $M^{2L_t}$ constellation points and has a
complexity $O(M^4)$ for $L_t=2$. \footnote{This statement applies
to the "exhaustive-search" ML demodulator; a triangular
decomposition of the channel matrix in itself, as in a standard
QRD, would be enough to lower the complexity of the search to
$O(M^{2L_t-1})$.}

The notation used in the sequel is that $\Omega_x$ will refer to
the {\it M}-PAM constellation for each real dimension.  Given the
formulation in (\ref{equivsys1})-(\ref{eq:utriangular}) and
neglecting scalar energy normalization factors to simplify the
notation, the ML decision metric becomes
\begin{eqnarray}
T(\textbf{x}_r)&=&\left\|\tilde{\textbf{y}}_r-\tilde{\textbf{R}}\textbf{x}_r\right\|^2=\frac{\left(\tilde{y}_1-\sigma^2_1x_{1}-s_1x_{3}-s_2x_{4}\right)^2}{\sigma^2_1}
+\frac{\left(\tilde{y}_2-\sigma^2_1x_{2}+s_2x_{3}-s_1x_{4}\right)^2}{\sigma^2_1}
\nonumber \\ &&
+\frac{\left(\tilde{y}_3-r_3x_{3}\right)^2}{\sigma^2_1r_3}
+\frac{\left(\tilde{y}_4-r_3x_{4}\right)^2}{\sigma^2_1r_3}
\label{LORDmetr2tx}
\end{eqnarray}
The ML demodulator finds the maximum value of the metric over all
possible values of the sequence $\textbf{x}_r$.  This search can
be greatly simplified by noting for given values of $x_{3}$ and
$x_{4}$ the maximum likelihood metric reduces to
\begin{equation}
T(\textbf{x}_r)=\frac{\left(\tilde{y}_1-\sigma^2_1x_{1}-C_1(x_3,x_4)\right)^2}{\sigma^2_1}
+\frac{\left(\tilde{y}_2-\sigma^2_1x_{2}-C_2(x_3,x_4)\right)^2}{\sigma^2_1}
+ C_3(x_3,x_4) \label{LORDmetr2tx_group}
\end{equation}
where
\begin{equation}
C_1(x_3,x_4)=s_1x_3+s_2x_4 \qquad C_2(x_3,x_4)=-s_2x_3+s_1x_4
\qquad C_3(x_3,x_4) \ge 0
\end{equation}
The originality of LORD stems from the fact that - as clear from
(\ref{LORDmetr2tx_group}) - the conditional ML decision on $x_1$
and $x_2$ can immediately be made by a simple threshold test,
i.e.,
\begin{equation} \label{eq:slice2tx}
\hat{x}_1(x_3,x_4)=\mbox{round}\left(\frac{\tilde{y}_1-C_1(x_3,x_4)}{\sigma^2_1}\right),
\quad
\hat{x}_2(x_3,x_4)=\mbox{round}\left(\frac{\tilde{y}_2-C_2(x_3,x_4)}{\sigma^2_1}\right).
\end{equation}
where the round operation is a simple slicing operation to the
constellation elements of $\Omega_x$. This property is direct
consequence of the orthogonality of the problem formulation. The
final ML  estimate is then given as
\begin{eqnarray}\label{eq:mlwd1}
&&
\left\{\hat{x}_1(\hat{x}_3,\hat{x}_4),\hat{x}_2(\hat{x}_3,\hat{x}_4),\hat{x}_3,\hat{x}_4\right\}=
\arg \stackrel{\textstyle \min}{\scriptstyle x_3,x_4 \in
\Omega_x^2}\left\{\frac{\left(\tilde{y}_1-\sigma^2_1\hat{x}_1(x_3,x_4)-C_1(x_3,x_4)\right)^2}{\sigma^2_1}
\nonumber\right. \\ &&
+\left.\frac{\left(\tilde{y}_2-\sigma^2_1\hat{x}_2(x_3,x_4)-C_2(x_3,x_4)\right)^2}{\sigma^2_1}
 + C_3(x_3,x_4) \right\}
\end{eqnarray}
This implies that the number of points that has to be searched in
this formulation to find the true ML estimator is $M^2$ (with two
slicing operations per searched point). This is a significant
saving in complexity.

It should be noticed that, in a direct analogy to
(\ref{shift_equivsys})-(\ref{shifteq}), the ML estimate could as
well be found minimizing the reordered ML decision metric
\begin{eqnarray}
T'(\textbf{x}_s)&=&\left\|\tilde{\textbf{y}}_s-\tilde{\textbf{R}}_s\textbf{x}_s\right\|^2=\frac{\left(\tilde{y}_{s1}-\sigma^2_2x_{3}-s_1x_{1}+s_2x_{2}\right)^2}{\sigma^2_2}
+\frac{\left(\tilde{y}_{s2}-\sigma^2_2x_{4}-s_2x_{1}-s_1x_{2}\right)^2}{\sigma^2_2}
\nonumber \\ &&
+\frac{\left(\tilde{y}_{s3}-r_3x_{1}\right)^2}{\sigma^2_2r_3}
+\frac{\left(\tilde{y}_{s4}-r_3x_{2}\right)^2}{\sigma^2_2r_3}.
\label{shiftedLORDmetr}
\end{eqnarray}
Similarly to (\ref{eq:slice2tx})-(\ref{eq:mlwd1}), minimization of
(\ref{shiftedLORDmetr}) can be accomplished considering all
possible $M^2$ values for $(x_1,x_2)$ and obtaining the
corresponding $\left(\hat{x}_3(x_1,x_2),\hat{x}_4(x_1,x_2)\right)$
through rounding operations to the constellation elements of
$\Omega_x$.

We observe that this reduced complexity ML demodulation is a
direct consequence of the reordered lattice formulation. Each
group of two rows in the model (\ref{eq:utriangular}) corresponds
to a transmit antenna, or {\it layer} (the two terms will be used
interchangeably in the remainder of the paper). Equation
(\ref{eq:slice2tx}) shows that the decisions for the top layer can
be made independently for the I and the Q modulation. If the
traditional lattice formulation \cite{DCB:00} is adopted instead,
in (\ref{LORDmetr2tx}) the partial ED (PED) terms corresponding to
the higher rows of the triangularized model become dependent on
all the lower layers of the transmit modulation, and the
simplified demodulation (\ref{eq:mlwd1}) is no longer possible.

Two further observations conclude this section:\begin{description}
    \item[-] The search of the lattice points can be carried out in a completely parallel fashion. This solves one of the drawbacks of SD algorithm, characterized by a recursive - i.e. serial - search, and is desirable for VLSI implementations.
    \item[-] The lattice point enumeration technique, i.e. method of spanning the points during the search, is not important for LORD as long as all $M^2$ possible cases for the bottom layer are searched. However, we observe that ordering the candidate list according to an increasing ED from the receiver observations has important implications for suboptimal
    searches, i.e. if less than $M^2$ values for the bottom layer are considered. This corresponds to the Schnorr-Euchner (SE) \cite{SE:94},\cite{EV:99} enumeration
method. Future work will address this important
    sub-optimal and reduced-complexity version of LORD.
\end{description}


\subsection{LLR generation} \label{sec_LLR_2TX}
This section deals with the reduced-complexity generation of
reliable soft output information. This problem is often neglected
in lattice decoding literature because of the intrinsic
difficulties caused by the SD attempt to reduce to the minimum the
number of searched lattice points. As mentioned in Section
\ref{sec_intro}, a partial solution to this issue has been
proposed in \cite{HTB:03}, \cite{BGF:03} with the introduction of
the so-called "candidate list". Unfortunately the random nature of
the selected points to be stored in this list pose several
implementation and complexity issues, also evidenced in
\cite{ZRF:04}. To name a few, no rule to optimally size the list
has been proposed, and simulation results show that in order to
obtain reliable LLRs the size depends on the considered MIMO
scenario; also, points are stored in the list in an inherently
sequential manner; the "quality" of the points stored in the list,
as well as the total number of searched sequences before the
search can be declared concluded, strongly depends on choice of
the sphere radius. The use of LR techniques can only help in
making the convergence to the ML solution faster, but precludes
the detection algorithm from computing soft-output values, because
the boundaries of the information set are no more recognizable
after the application of such techniques. The choice followed in
this work was then to avoid LR methods and to solve the
indeterministic and sequential nature of the selection of the
sequences needed for the generation of reliable bit soft-output
information.

 The problem is first recalled for the complex-domain
system (\ref{cmplx_sys_eq}). If $M^2$-QAM constellation is
considered for the information symbol vector and $M_c$ is the
number of bits per symbol, the LLR or logarithmic a-posteriori
probability (APP) ratio of the bit $b_k$, $k=1, \ldots , 2M_c\,$,
conditioned on the received channel symbol vector $\textbf{y}_c$,
is often expressed as:
\begin{equation}
L(b_k|\textbf{y}_c)=\ln{\frac{P(b_k=1|\textbf{y}_c)}{P(b_k=0|\textbf{y}_c)}}=\ln{\frac{\displaystyle\sum_{{\bf
x}_c \in
S(k)^+}P(\textbf{y}_c|\textbf{x}_c)P_a(\textbf{x}_c)}{\displaystyle\sum_{{\bf
x}_c \in S(k)^-}P(\textbf{y}_c|\textbf{x}_c)P_a(\textbf{x}_c)}}
\label{generalLLR}
\end{equation}
In (\ref{generalLLR}) $S(k)^+$ ($S(k)^-$) is the set of
$2^{2Mc-1}$ bit sequences having $b_k=1$ ($b_k=0$); $P_a({\bf
x}_c)$ represents the a-priori probabilities of $\textbf{x}_c$ and
will be neglected in the rest of this paper as equiprobable
transmit symbols are considered. From (\ref{cmplx_sys_eq}), the
likelihood function $\displaystyle P(\textbf{y}_c|\textbf{x}_c)$
is given by:
\begin{equation}
P(\textbf{y}_c|\textbf{x}_c)\propto
\exp{\left[-\frac{1}{2\sigma^2}\|\textbf{y}_c-\sqrt{\frac{E_s}{2}}\textbf{Hx}_c\|^2
\right]}=\exp{\left[-D(\textbf{x}_c)\right]}
\end{equation}
where $\sigma^2=N_0/2$ and $D(\textbf{x}_c)$ is the ED term. The
summation of exponentials involved in (\ref{generalLLR}) can be
approximated according to the following so-called max-log
approximation:
\begin{equation}
\ln{\sum_{{\bf x}\in S(k)^+} \exp{\left[-D({\bf
x})\right]}}\approx\ln{\max_{{\bf x}\in S(k)^+}{\exp{\left[-D({\bf
x})\right]}}}=-\min_{{\bf x}\in S(k)^+}{D({\bf x})} \label{max}
\end{equation}
Expression (\ref{max}) is equivalent to neglecting a correction term
in the exact log-domain version of (\ref{generalLLR}), which
uses the ``Jacobian logarithm'' or $\max^*$ function
\begin{equation}
\textrm{jacln}(a,b):=\ln{\left[\exp{(a)}+\exp{(b)}\right]}=\max{(a,b)}+\ln{\left[1+\exp{(-|a-b|)}\right]}.
\label{maxstar}
\end{equation}
As shown e.g. in \cite{RVH:95}, the performance degradation caused
by the max-log approximation is generally very small compared to
the use of the $\max^*$ function. Using (\ref{max}) in
(\ref{generalLLR}), max-log bit LLRs can then be written as:
\begin{equation}
L(b_k|\textbf{y}_c)\approx\min_{{\bf x}_c\in
S(k)^-}{D(\textbf{x}_c)}-\min_{{\bf x}_c\in
S(k)^+}{D(\textbf{x}_c)} \label{maxlogLLR}
\end{equation}
Expression (\ref{maxlogLLR}) involves two minimization problems,
i.e. for each bit index $k=1, \ldots , 2M_c\,$ it requires
identification of the most likely transmit sequence (or lattice
point) where $b_k=1$ and the most likely one where $b_k=0$. By
definition, one of the two sequences is the hard-decision ML
solution of (\ref{MLd}). However, using SD, there is no guarantee
that the other sequence is found during the lattice search. LORD
does not have this problem, as shown in the sequel.

The formulation of the problem in case of real-domain lattice
equations is perfectly similar. From (\ref{equivsys1}) the LLRs
assume the form:
\begin{equation}
L(b_k|\tilde{\textbf{y}}_r)=\ln{\frac{\displaystyle\sum_{{\bf x}_r
\in
S(k)^+}P(\tilde{\textbf{y}}_r|\textbf{x}_r)}{\displaystyle\sum_{{\bf
x}_r \in S(k)^-}P(\tilde{\textbf{y}}_r|\textbf{x}_r)}}
\label{realLLR}
\end{equation}
where, recalling (\ref{LORDmetr2tx}), the likelihood function
$\displaystyle P(\tilde{\textbf{y}}_r|\textbf{x}_r)$ is given by:
\begin{equation}\label{lklfctn}
P(\tilde{\textbf{y}}_r|\textbf{x}_r)=\exp\left[-|T(\textbf{x}_r)|\right].
\end{equation}
Let us first focus on the bits corresponding to the complex symbol
$X_{c2}$ in the symbol sequence $\textbf{x}_c=(X_{c1} \, \,
X_{c2})^{T}$. By employing arguments similar to those that led to
the simplified ML estimator (\ref{eq:mlwd1}), it can be easily
proven that the two ED terms needed for every bit in $X_{c2}$ are
certainly found computing (\ref{LORDmetr2tx}) over the possible
$M^2$ values of $X_{c2}=(x_3, \, x_4)$ and minimizing the
expression over $X_{c1}=(x_1, \, x_2)$, for every value of
$X_{c2}$. This last operation is simply carried out through the
slicing operation to the constellation elements of $\Omega_x$,
described in (\ref{eq:slice2tx}). The LLRs relative to the bits
corresponding to $X_2$, $b_{2,k}$, can then be written as:
\begin{equation}
L(b_{2,k}|\tilde{\textbf{y}})\approx\min_{x_3,x_4\in
S(k)_2^-}{T(\textbf{x}_r)}-\min_{x_3,x_4\in
S(k)_2^+}{T(\textbf{x}_r)} \label{LLR_X2}
\end{equation}
where $k=1, \ldots , M_c$, and $S(k)_2^+$ ($S(k)_2^-$) are the set
of $2^{M_c-1}$ bit sequences having $b_{2,k}=1$ ($b_{2,k}=0$).

The computation of the LLRs for the bits corresponding to symbol
$X_1$ can be obtain by a simple reordering of the model and a
repeating of the LORD processing, as for
(\ref{shift_equivsys})-(\ref{shifteq}). Recalling that
$\textbf{x}_s=\left[x_3 \, x_4 \, x_1 \, x_2 \right]$ is the
reordered information sequence, using (\ref{shiftedLORDmetr}) the
LLRs of the bits corresponding to $X_1$, $b_{1,k}$, can be written
as
\begin{equation}
L(b_{1,k}|\tilde{\textbf{y}_s})\approx\min_{x_1,x_2\in
S(k)_1^-}{T'(\textbf{x}_s)}-\min_{x_1,x_2\in
S(k)_1^+}{T'(\textbf{x}_s)} \label{LLR_X1}
\end{equation}
where $k=1, \ldots , M_c$, $S(k)_1^+$ ($S(k)_1^-$) are the set of
$2^{M_c-1}$ bit sequences having $b_{1,k}=1$ ($b_{1,k}=0$). There
is significant complexity reduction that can be utilized in
forming the LLR. By comparing (\ref{eq:utriangular}) and
(\ref{shifteq}), it is apparent that much of the preprocessing
computation needed in (\ref{eq:utriangular}) can be used in the
reordered (\ref{shifteq}). The resulting complexity of the
preprocessing stage will be $O(16L_r+12)$. The lattice search for
both orderings will have complexity $O(2M^2)$ due to the max-log
LLR computation.

\section{LORD Algorithm - case of $L_t$ transmit antennas} \label{sec_LORD_NTX}
\begin{figure}
\centering
\includegraphics[width=6.5in]{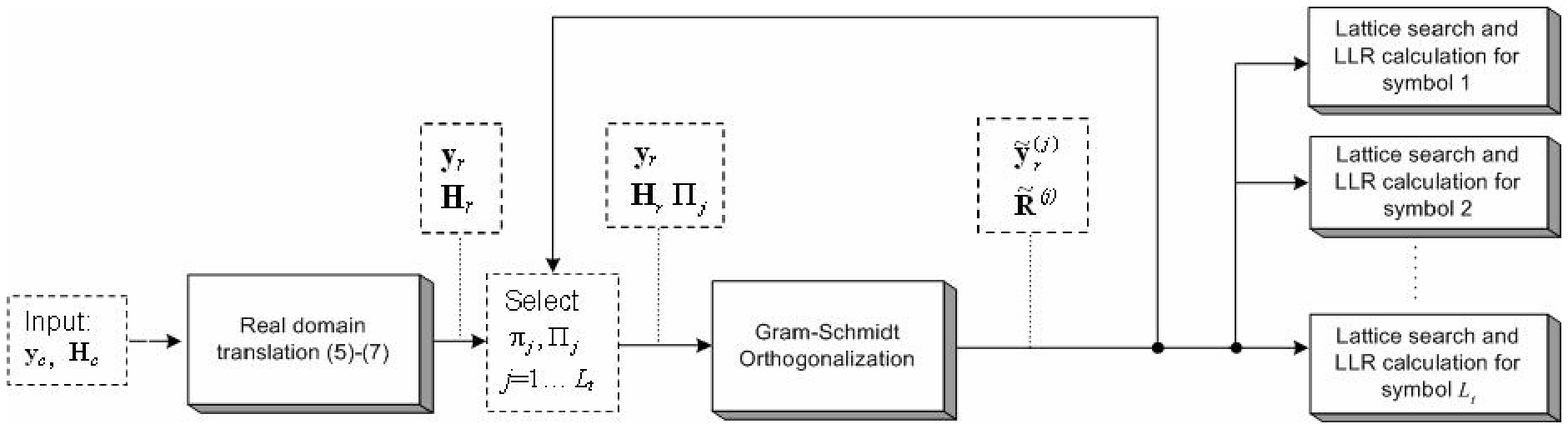}
\caption{Block diagram of LORD detector.}
\label{fig:LORD_blk_diag}
\end{figure}

The LORD detection algorithm can be generalized to any $L_t>2$ and
$L_r \geq L_t$ in a sub-optimal way but still often remaining
near-ML, as shown in Section \ref{sec_performance}. Specifically,
a computationally efficient QRD algorithm is described in Section
\ref{sec_preproc_NTX1}. A notationally compact and elegant
recursive variant of QRD is given in the Appendix. The relation
between the extended and the compact representations is analogous
to that existing between QRD through GSO and modified GSO (MGSO)
\cite{Golub}. The main difference between the GSO proposed in this
paper and the QRD \cite{Golub} is represented by the way the
normalizations are handled, as clear from the sequel. The lattice
search and soft output generation are then obtained generalizing
the steps described in \ref{sec_latsearch_2TX} and
\ref{sec_LLR_2TX} respectively. A block diagram highlighting LORD
algorithm steps is reported in Fig. \ref{fig:LORD_blk_diag}.

\subsection{The preprocessing algorithm - standard formulation} \label{sec_preproc_NTX1}

 The formulation described in the sequel can be viewed as a generalization of the
equations reported in Section \ref{sec_preproc_2TX} for $L_t=2$.
This preprocessing corresponds to GSO with normalizations deferred
to a later stage. To best understand this preprocessing note that
there is an $2L_t \times 2L_r$ orthogonal matrix
\begin{equation}
{\bf Q}=\left[\begin{array}{ccc} \textbf{q}_1 \quad \textbf{q}_2
\quad \textbf{q}_3 \quad \textbf{q}_4 \, \ldots \,
\textbf{q}_{2L_t-1} \quad \textbf{q}_{2L_t} \end{array} \right].
\end{equation}
where
\begin{eqnarray} \label{q_expr}
\textbf{q}_1 &=&\textbf{h}_1 \\
\textbf{q}_2 &=&\textbf{h}_2 \nonumber \\
\textbf{q}_3 &=&
\sigma_1^2\textbf{h}_3-s_{1,3}\textbf{h}_1- s_{2,3}\textbf{h}_2
\nonumber \\
\textbf{q}_4 &=&
\sigma_1^2\textbf{h}_4-s_{1,4}\textbf{h}_1- s_{2,4}\textbf{h}_2
\nonumber \\ \textbf{q}_5 &=&
r_3\sigma_1^2\textbf{h}_5-r_3s_{1,5}\textbf{h}_1-
r_3s_{2,5}\textbf{h}_2-t_{3,5}\textbf{q}_3-t_{4,5}\textbf{q}_4\\
\vdots \nonumber \\
\textbf{q}_{p}
&=&P_1^k\left[\sigma_1^2\textbf{h}_{p}-s_{1,p}\textbf{h}_1-
s_{2,p}\textbf{h}_2\right]-\sum_{i=2}^{k-1}\left[P_{i+1}^k\left(t_{2i-1,p}\textbf{q}_{2i-1}+
t_{2i,p}\textbf{q}_{2i}\right)\right]-t_{2k-1,p}\textbf{q}_{2k-1}-t_{2k,p}\textbf{q}_{2k}
 \nonumber
\end{eqnarray}
where $p$ denotes the generic $k-th$ pair of \textbf{q} columns,
i.e. $p=\{2k+1,2k+2\}$, with $k=\{2,\ldots ,L_t-1\}$, and which
uses the following definitions:
\begin{equation}
s_{j,k}\equiv\textbf{h}_j^T\textbf{h}_k, \qquad
t_{j,k}\equiv\textbf{q}_j^T\textbf{h}_k, \qquad
\sigma_k^2\equiv\left\|h_k^2\right\| \qquad
P_m^n\equiv \prod_{j=m}^{n}r_{2j-1}
\end{equation}
where $m,n$ are integers with $1\leq m \leq n$.
The terms
$r_{2k-1}$, with $k=\{1,\ldots L_t\}$, are given by:
\begin{eqnarray} \label{r_expr}
r_1&=&1 \nonumber \\ \quad r_3&=&\quad\sigma_3^2 \sigma_1^2
-s_{1,3}^2-s_{2,3}^2
\\ \vdots \nonumber \\
r_{2k-1}&=&P_2^{k-1}\left(\sigma_1^2\sigma_{2k-1}^2-s_{1,2k-1}^2-s_{2,2k-1}^2\right)-
\sum_{i=2}^{k-2}{P_{i+1}^{k-1}\left(t_{2i-1,2k-1}^2+t_{2i,2k-1}^2\right)}\\
&&-t_{2k-3,2k-1}^2-t_{2k-2,2k-1}^2. \nonumber
\end{eqnarray}
They can also be written in the compact form
\begin{equation}
r_{2k-1}=P_2^{k-1}\sigma_1^2\sigma_{2k-1}^2(1-|\rho_{1,k}|^2
-\sum_{i=2}^{k-1}|{\rho\prime}_{i,k}|^2) \label{r_expr:2}
\end{equation}
 where we have used the square magnitudes of the (generalized) correlation coefficients:
\begin{eqnarray} \label{rho_expr}
|\rho_{k,j}|^2&=&\frac{s_{2k-1,2j-1}^2+
s_{2k-1,2j}^2}{\sigma_{2k-1}^2\sigma_{2j-1}^2}\\
|{\rho\prime}_{k,j}|^2&=&\frac{t_{2k-1,2j-1}^2+
t_{2k-1,2j}^2}{\left\|\textbf{q}_{2k-1}\right\|^2\sigma_{2j-1}^2},
\, j>k. \nonumber
\end{eqnarray}
It is easily shown that (\ref{h_norms}) can be generalized as:
\begin{eqnarray}\label{q_norms}
\|\textbf{q}_{2k-1}\|^2&=&\|\textbf{q}_{2k}\|^2=P_1^k \sigma_1^2
\\ \textbf{q}_{2k-1}^T\textbf{h}_{2j-1}&=&\textbf{q}_{2k}^T\textbf{h}_{2j},
\qquad
\textbf{q}_{2k-1}^T\textbf{h}_{2j}=-\textbf{q}_{2k}^T\textbf{h}_{2j-1},
\, j>k. \nonumber
\end{eqnarray}
Also, by construction the \textbf{q} vectors, and
\{\textbf{q},\textbf{h}\} couples, are pairwise orthogonal, i.e.
\begin{equation}
\nonumber \textbf{q}_{2k-1}^T\textbf{q}_{2k}=0, \quad
\textbf{q}_{2k-1}^T\textbf{h}_{2k}=0
\end{equation}
The orthogonal matrix \textbf{Q} then satisfies
\begin{equation}
{\bf Q}^T{\bf Q}=\mbox{diag}\left[\sigma_1^2, \ \sigma_1^2, \
\left\|\textbf{q}_3\right\|^2, \ \left\|\textbf{q}_3\right\|^2,
\ldots , \ \left\|\textbf{q}_{2L_t-1}\right\|^2, \
\left\|\textbf{q}_{2L_t-1}\right\|^2 \right]
\end{equation}
By defining the $2L_t \times 2L_t$ upper triangular matrix
\begin{equation}
{\bf R}=\left[\begin{array}{cccccccccc}
1 & 0 & s_{1,3} & s_{1,4} & r_3 s_{1,5} & \ldots & \ldots & \ldots & P^{L_t-1}_1 s_{1,2L_t-1} & P^{L_t-1}_1 s_{1,2L_t}\\
0 & 1 & -s_{1,4} & s_{1,3} & -r_3 s_{1,6} & \ldots & \ldots & \ldots & -P^{L_t-1}_1 s_{1,2L_t} & P^{L_t-1}_1 s_{1,2L_t-1}\\
0 & 0 & 1 & 0 & t_{3,5} & \ldots & \ldots & \ldots & P^{L_t-1}_2 t_{3,2L_t-1} & P^{L_t-1}_2 t_{3,2L_t} \\
0 & 0 & 0 & 1 & t_{4,5} & \ldots & \ldots & \ldots & -P^{L_t-1}_2 t_{3,2L_t} & P^{L_t-1}_2 t_{3,2L_t-1} \\
\ldots & \ldots & \ldots & \ldots & \ldots & \ldots &\ldots &\ldots &\ldots &\ldots \\
0 & 0 & 0 & 0 & 0 & \ldots &1 &0 & t_{2L_t-3,2L_t-1} & t_{2L_t-3,2L_t} \\
0 & 0 & 0 & 0 & 0 & \ldots &0 &1 &-t_{2L_t-3,2L_t} & t_{2L_t-3,2L_t-1} \\
0 & 0 & 0 & 0 & 0 & \ldots &0 &0 & 1 &0 \\
0 & 0 & 0 & 0 & 0 & \ldots &0 &0 & 0 &1 \\
\end{array}\right]
\end{equation}
the real channel matrix $\textbf{H}_r$ can be decomposed in the
product:
\begin{equation}
{\bf H}_r={\bf Q}{\bf R}{\bf \Lambda}_q
\end{equation}
where the $2L_t \times 2L_t$ diagonal matrix
\begin{equation}
{\bf \Lambda}_q=\mbox{diag}\left[1, 1, \ \sigma_1^{-2},
\sigma_1^{-2}, \ldots
\left(P_1^{L_t-1}\sigma_1^2\right)^{-1}\right]
\end{equation}
includes the normalization factors due to the fact that \textbf{Q}
is not orthonormal. Note again all values of ${\bf Q}$ are simple
functions of the known channel coefficients. Again the signal model after
preprocessing is given as
\begin{equation}
\tilde{\textbf{y}}_r=\sqrt{\frac{E_{s}}{2}}
\, {\bf \tilde{R}} \textbf{x}_r + {\bf Q}^T
\textbf{n}_r=\sqrt{\frac{E_{s}}{2}} \, \tilde{{\bf R}}
\textbf{x}_r + \tilde{\textbf{n}}_r. \label{equivsys}
\end{equation}
The triangular matrix $\tilde{{\bf R}}={\bf
Q}^T{\bf Q}{\bf R}{\bf \Lambda}_q$, given by:
\begin{equation} \label{eq:rtildeNtx}
\tilde{{\bf R}}=\left[\begin{array}{cccccccccc}
\sigma_1^2 & 0 & s_{1,3} & s_{1,4} & s_{1,5} & \ldots & \ldots & \ldots & s_{1,2L_t-1} & s_{1,2L_t}\\
0 & \sigma_1^2 & -s_{1,4} & s_{1,3} & -s_{1,6} & \ldots & \ldots & \ldots & -s_{1,2L_t} & s_{1,2L_t-1}\\
0 & 0 & r_3 & 0 & t_{3,5} & \ldots & \ldots & \ldots & t_{3,2L_t-1} & t_{3,2L_t} \\
0 & 0 & 0 & r_3 & -t_{3,6} & \ldots & \ldots & \ldots & -t_{3,2L_t} & t_{3,2L_t-1} \\
\ldots & \ldots & \ldots & \ldots & \ldots & \ldots &\ldots &\ldots &\ldots &\ldots \\
0 & 0 & 0 & 0 & 0 & \ldots &r_{2L_t-3} &0 & t_{2L_t-3,2L_t-1} & t_{2L_t-3,2L_t} \\
0 & 0 & 0 & 0 & 0 & \ldots &0 &r_{2L_t-3} &-t_{2L_t-3,2L_t} & t_{2L_t-3,2L_t-1} \\
0 & 0 & 0 & 0 & 0 & \ldots &0 &0 & r_{2Lt-1} &0 \\
0 & 0 & 0 & 0 & 0 & \ldots &0 &0 & 0 &r_{2Lt-1} \\
\end{array}\right]
\end{equation}
The noise vector in the triangular model still has independent
components but with unequal variances given by\footnote{It should
be noted that in practical implementations the normalizations
should be performed just prior to the lattice search in order
avoid including the different noise variances in the ED
computation (\ref{LORDmetrNtx}).}:
\begin{eqnarray}
{\bf
R}_{\tilde{n}}=E\left[\tilde{\textbf{n}}\tilde{\textbf{n}}^T\right]
=\frac{N_0}{2}
\mbox{diag}\left[\sigma_1^2, \ \sigma_1^2, \ \ldots, \
P_1^{L_t}\sigma_1^2, \ P_1^{L_t}\sigma_1^2 \right]. \nonumber
\end{eqnarray}
The resulting preprocessing complexity expressed in terms of RMs
is $O(2L_rL_t^2+2L_t^2+4L_tL_r+K)$, where $K=13$ for $L_t=4$ and
grows asymptotically as $\frac{21}{2}L_t^2$ for large $L_t$. More
detailed explanations on complexity are reported in Section
\ref{sec_complexity}. We note that this result takes into account
that an explicit computation of the matrix \textbf{Q} is not
required, but rather it is possible to proceed to the direct
computation of the scalar products $\textbf{Q}^T\textbf{y}_r$.
Also, the benefit of deferring the normalizations will become
apparent from Sections \ref{sec_LLR_NTX} and \ref{sec_complexity}.

\subsection{Lattice search and demodulation} \label{sec_latsearch_NTX}
Having the matrix $\textbf{Q}$ allows an observation model like
(\ref{equivsys}) to be derived and a simplified demodulation is possible.
 Using the structure of
$\tilde{\textbf{R}}$ shown in (\ref{eq:rtildeNtx}), the decision metrics
can be written as:
\begin{eqnarray} \label{LORDmetrNtx}
T(\textbf{x}_r)=\left\|\tilde{\textbf{y}}_r-\tilde{\textbf{R}}\textbf{x}_r\right\|^2&=&\frac{\left(\tilde{y}_1-\sigma^2_1x_{1}-\sum_{k=3}^{2L_t}s_{1,k}x_k\right)^2}{\sigma^2_1}
\\ && +
\frac{\left(\tilde{y}_2-\sigma^2_1x_{2}-\sum_{k=3}^{2L_t}s_{2,k}x_k\right)^2}{\sigma^2_1}
\nonumber \\ &&
+\frac{\left(\tilde{y}_3-r_3x_{3}-\sum_{k=5}^{2L_t}t_{3,k}x_k\right)^2}{\sigma^2_1r_3}
+ \ldots \nonumber \\ && +
\frac{\left(\tilde{y}_{2L_t-1}-r_{2L_t-1}x_{2L_t-1}\right)^2+\left(\tilde{y}_{2L_t}-r_{2L_t-1}x_{2L_t}\right)^2}{\sigma^2_1P_2^{L_t-1}}
\nonumber
\end{eqnarray}
The proposed simplified demodulation consists of considering all
$M^2$ values for the I and Q couples of the lowest level layer.
For each hypothesized value of $x_{2L_t-1}$ and $x_{2L_t}$, here
denoted $\tilde{x}_{2L_t-1}$ and $\tilde{x}_{2L_t}$, the higher
level layers are decoded through interference nulling and
cancelling, or ZF-DFE. For a given layer ordering these operations
are similar to the QR version of the O-DFE algorithm except for
the important difference represented by operating in the real
domain through a novel lattice formulation. The estimation of the
I and Q of the remaining $L_t-1$ symbols is implemented through a
slicing operation to the constellation elements of $\Omega_x$ for
$x_1,\dots x_{2L_t-2}$. By writing:
\begin{eqnarray}
T(\textbf{x}_r)&=&\frac{\left(\tilde{y}_1-\sigma^2_1x_{1}-C_1\left(x_3,
\dots x_{2L_t}\right)\right)^2}{\sigma^2_1} \nonumber \\ && +
\frac{\left(\tilde{y}_2-\sigma^2_1x_{2}-C_2\left(x_3, \dots
x_{2L_t}\right)\right)^2}{\sigma^2_1} \nonumber \\ &&
+\frac{\left(\tilde{y}_3-r_3x_{3}-C_3\left(x_5, \dots
x_{2L_t}\right)\right)^2}{\sigma^2_1r_3} \nonumber \\ && + \ldots
+ C_{2L_t-1}\left(x_{2L_t-1},x_{2L_t}\right)
\label{LORDmetrNtxgroup}
\end{eqnarray}
where
\begin{equation}\label{lastterm}
C_{2L_t-1}=\frac{\left(\tilde{y}_{2L_t-1}-r_{2L_t-1}x_{2L_t-1}\right)^2+\left(\tilde{y}_{2L_t}-r_{2L_t-1}x_{2L_t}\right)^2}{\sigma^2_1P_2^{L_t-1}}
\end{equation}
then the conditionally decoded values of $x_1,\dots x_{2L_t-2}$ as
function of each candidate couple
$\left(\tilde{x}_{2L_t-1},\tilde{x}_{2L_t}\right)$ are determined
recursively as:
\begin{eqnarray} \label{eq:slice}
\hat{x}_{2L_t-2}&=&\mbox{round}\left(\frac{\tilde{y}_{2L_t-2}-
C_{2L_t-2}\left(\tilde{x}_{2L_t-1},\tilde{x}_{2L_t}\right)}{r_{2L_t-3}}\right)
\nonumber\\&& \vdots \nonumber \\
\hat{x}_1&=&\mbox{round}\left(\frac{\tilde{y}_1-C_1\left(\hat{x}_3,
\dots,
\hat{x}_{2L_t-2},\tilde{x}_{2L_t-1},\tilde{x}_{2L_t}\right)}{\sigma^2_1}\right)
\end{eqnarray}
Denoting these $2L_t-2$ conditional decisions as $\hat{\textbf{x}}^{(-)}
\left(\tilde{x}_{2L_t-1},\tilde{x}_{2L_t}\right)$, the resulting sequence
estimate is then determined as:
\begin{equation}\label{eq:mlwd_gen}
\hat{\textbf{x}}_r=\left\{\hat{\textbf{x}}^{(-)}\left(\hat{x}_{2L_t-1},\hat{x}_{2L_t}\right),
\hat{x}_{2L_t-1},\hat{x}_{2L_t}\right\}
\end{equation}
where
\begin{equation}
\left\{\hat{x}_{2L_t-1},\hat{x}_{2L_t}\right\} = \arg
\min_{\tilde{x}_{2L_t-1},\tilde{x}_{2L_t}
\in \Omega_x^2} T(\hat{\textbf{x}}^{(-)}
\left(\tilde{x}_{2L_t-1},\tilde{x}_{2L_t}\right),\tilde{x}_{2L_t-1},
\tilde{x}_{2L_t} )
\end{equation}
Recall each group of two rows of $\tilde{{\bf R}}$ in
(\ref{eq:rtildeNtx}) corresponds to a transmit antenna. At the
bottom of the triangularized model the search for the I and Q of
the $L_t$-th transmit antenna is broken into orthogonal dimensions
and can be carried out independently. Also, looking at each $k$-th
pair of rows ($2k-1,2k$) of (\ref{eq:rtildeNtx}) it is clear that
the corresponding I and Q couple ($x_{2k-1},x_{2k}$) can be
decoded independently once the interference from the lower layers
has been cancelled. These orthogonality relations were not true
for the traditional lattice formulation \cite{DCB:00}. Differently
from the case of $L_t=2$ transmit antennas, however, the
generalized low-complexity search is suboptimal. A lower
complexity optimal ML demodulation would still be possible through
slicing ($x_1,x_2$) over all the possible $M^{2(L_t-1)}$ values of
the other elements, but this would still be too complex for
$L_t>2$. Near-optimal hard-output performance would be possible if
the layers are ordered properly in the above described
demodulation scheme, as it will be highlighted in future works.
Simulation results, not reported in the present paper, confirm
this statement. The next section will show through numerical
results that ordering is not essential in order to achieve near-ML
performance in BICM systems.

\subsection{Bit LLR generation} \label{sec_LLR_NTX}

The proposed idea is to approximate the minimization of the two
terms involved in (\ref{maxlogLLR}) using the principles
exemplified with (\ref{LORDmetrNtx}-\ref{eq:mlwd_gen}). Let us
consider the bits corresponding to the complex symbol $X_{L_t}$ in
the symbol vector $\textbf{x}_c=(X_1,\ldots X_{L_t})^{T}$. The
sequences used to minimize the two terms of (\ref{maxlogLLR}) are
determined considering all possible $M^2$ values for $X_{L_t}$,
while the value for the other elements ($x_1,\ldots x_{2L_t-2}$)
is derived through the DFE operation (\ref{eq:slice}). Equation
(\ref{maxlogLLR}) can then be approximated as:
\begin{eqnarray}
L(b_{L_t,k}|\tilde{\textbf{y}})&=&\min_{\left\{\tilde{x}_{2L_t-1},
\tilde{x}_{2L_t}\right\}
\in S(k)_{L_t}^-} T(\hat{\textbf{x}}^{(-)}
\left(\tilde{x}_{2L_t-1},\tilde{x}_{2L_t}\right),\tilde{x}_{2L_t-1},
\tilde{x}_{2L_t} ) \nonumber \\
&&-\min_{\left\{\tilde{x}_{2L_t-1}, \tilde{x}_{2L_t}\right\}\in
S(k)_{L_t}^+} T(\hat{\textbf{x}}^{(-)}
\left(\tilde{x}_{2L_t-1},\tilde{x}_{2L_t}\right),\tilde{x}_{2L_t-1},
\tilde{x}_{2L_t} )
\label{LLR_X}
\end{eqnarray}
where $b_{L_t,k}$ are the bits corresponding to $X_{L_t}$, $k=0,
\ldots , M_c-1$, and $S(k)_{L_t}^+$ $\left(S(k)_{L_t}^-\right)$
are the set of $2^{M_c-1}$ complex symbols having $b_{L_t,k}=1$
($b_{L_t,k}=0$).

In order to compute the approximated max-log LLRs also for the
bits corresponding to the other $L_t-1$ symbols in $\textbf{x}_c$,
the algorithm has to compute the steps formerly described for
different layer orderings, where in turn each layer becomes the
reference one only once. In other words, we need models where the
last two rows of the triangular matrix (\ref{eq:rtildeNtx})
correspond, in turn, to every symbol in $\textbf{x}_c$. This can
be accomplished starting from the natural integer order sequence
$\textbf{x}_c$ and generating the other $L_t-1$ permutations
recursively by exchanging the last layer with all the others;
then, the columns of the real channel matrix $\textbf{H}_r$ have
to be permuted accordingly, prior to performing the GSO.

Some considerations on the resulting preprocessing complexity are
in order here. The overall complexity can be estimated recalling
that by applying the GSO, the QRD computes the matrix
$\tilde{\bf{R}}$ line by line from top to bottom and the matrix
${\bf Q}$ columnwise from left to right, as clear from
(\ref{q_expr}) and (\ref{r_expr}). This would suggest that in
order to minimize the complexity the considered permutations
should differ for the least possible number of indexes. In this
case many operations would not have to be recomputed for different
symbol orderings. Anyway, the core of the processing consisting in
the scalar products between $2L_r$-element vectors can be computed
only once thus keeping an overall cubic complexity with the number
of antennas. This is a consequence of the absence of
normalizations in the GSO computation, as better detailed in
Section \ref{sec_complexity}.

For the sake of argument, let us consider the following set of
index permutations of the complex symbol sequence $\textbf{x}_c$.
Let $\pi_{L_t}$ be the natural integer order index set, where the
reference layer is the $L_t$-th. Then, a possible efficient set for a
recursive APP computation is:
\begin{eqnarray}
\label{perms} \pi_{L_t}&=&{1,...L_t} \\
\pi_{L_t-1}&=&{1,\ldots L_{t-2}, L_t, L_{t-1}} \nonumber \\
\pi_{L_t-2}&=&{1,\ldots L_{t-1}, L_t, L_{t-2}} \nonumber \\
&& \vdots \nonumber \\
\pi_1&=&{2, 3, \ldots L_t, 1} \nonumber
\end{eqnarray}
Let $\Pi_j$ denote a $2L_t \times 2L_t$ permutation matrix such
that arranges the columns of $\textbf{H}_r$ according to the index
set $\pi_j$. Then the GSO yields:
\begin{equation}
{\bf H}_r\Pi_j={\bf Q}^{(j)}{\bf R}^{(j)}{\bf \Lambda}_q^{(j)}
\end{equation}
and the matrix $\tilde{\bf{R}}^{(j)}$ can be computed as
$\tilde{{\bf R}}^{(j)}={\bf Q}^{(j)\,T}{\bf Q}^{(j)}{\bf
R}^{(j)}{\bf \Lambda}_q^{(j)}$. Finally, we can write:
\begin{equation}
\tilde{\textbf{y}}_r^{(j)}={\bf Q}^{(j)\,T}
\textbf{y}_r=\sqrt{\frac{E_{s}}{L_t}} \, {\bf \tilde{R}}^{(j)}
\textbf{x}_r^{(j)} + {\bf Q}^{(j)\,T} \textbf{n}_r
\end{equation}
where $\textbf{x}_r^{(j)}$ is the permuted I and Q sequence.
Indicating the corresponding ED metrics as
$T^{(j)}(\textbf{x}_r^{(j)})$, the LLR of the bits corresponding
to the $j$-th symbol can be written as:
\begin{eqnarray}
L(b_{j,k}|\tilde{\textbf{y}}^{(j)})&=&\min_{\left\{\tilde{x}_{2j-1},
\tilde{x}_{2j}\right\} \in S(k)_{j}^-}
T^{(j)}(\hat{\textbf{x}}_j^{(-)}
\left(\tilde{x}_{2j-1},\tilde{x}_{2j}\right),\tilde{x}_{2j-1},
\tilde{x}_{2j} ) \nonumber \\
&&-\min_{\left\{\tilde{x}_{2j-1}, \tilde{x}_{2j}\right\}\in
S(k)_{j}^+} T^{(j)}(\hat{\textbf{x}}_j^{(-)}
\left(\tilde{x}_{2j-1},\tilde{x}_{2j}\right),\tilde{x}_{2j-1},
\tilde{x}_{2j} ) \label{LLR_X_j}
\end{eqnarray}
where $b_{j,k}$ are the bits corresponding to $X_{j}$, $k=0,
\ldots , M_c-1$, $S(k)_{j}^+$ ($S(k)_{j}^-$) are the set of
$2^{M_c-1}$ bit sequences having $b_{j,k}=1$ ($b_{j,k}=0$), and
$\textbf{x}_j^{(-)}\left(\tilde{x}_{2j-1},\tilde{x}_{2j}\right)$
denotes the $2L_t-2$ conditional decisions of the layer order
sequence $\pi_j$ in (\ref{perms}), in analogy to (\ref{eq:slice}).

It is apparent that LORD is an approximated method for bit LLR
generation relying on a lattice search of $L_t M^2$ symbol
sequences as opposed to a search of $M^{2L_t}$ as required by the
maximum a-posteriori probability (MAP) demodulator. A further
practical advantage of LORD is that the LLR computation for the
bits corresponding to the $L_t$ symbols can be carried out in a
parallel fashion.

\subsection{Complexity estimation} \label{sec_complexity}

The aim of this section is to clarify the complexity estimates
previously reported, focusing on the general case of a
single-carrier MIMO system with $L_t$ transmit and $L_r$ receive
antennas and known CSI at the receiver. The estimates are
expressed in terms of RMs. For static or slowly-varying channel
applications, like WLANs, it is also important to distinguish
between channel-dependent and receiver observation-dependent
terms, because in this case CSI can be computed once per frame (or
packet) differently from the observation-related terms.

\noindent$\bullet$ \textit{Channel dependent terms.}

\noindent They are the entries of the matrix $\tilde{\textbf{R}}$
in (\ref{eq:rtildeNtx}). A significant observation is that the
number of nonzero real entries to be computed is $L_t^2$, instead
of $2L_t^2+L_t$. This is a consequence of the adopted I and Q
ordering and particularly of (\ref{h_norms}), (\ref{q_norms}).
\begin{description}
\item[-] Each of the $L_t^2$ entries involves the computation of
the scalar product of a $2L_r$-element vector, for a resulting
complexity $O(2L_rL_t^2)$. Specifically, they are
$\sigma_{2k-1}^2=\left\|h_{2k-1}^2\right\|$, with $k=1, \ldots
L_t$, and the terms $s_{i,j}=\textbf{h}_i^T\textbf{h}_j$, with
$i<j$ and $j=1, \ldots 2L_t$; it should be noticed that also
$t_{i,j}=\textbf{q}_i^T\textbf{h}_j$ ultimately depend on
$s_{i,j}$, as clear from (\ref{q_expr}).

\item[-] The computation of the terms $t_{i,j}$ grows
quadratically with $L_t$ for $L_t\geq4$, when $L_t-2$ couple of
columns of $\tilde{\textbf{R}}$ including those terms are present.
When $L_t=2$ no such terms exist, while there are only terms
$t_{i,j}$ involving $\textbf{q}_3$,$\textbf{q}_4$ if $L_t=3$, i.e.
the $t_{i,j}$ do not depend recursively on themselves as evident
from (\ref{q_expr}). The complexity associated with these
computations is then
\begin{eqnarray}
\label{K_expr}
K&=&6, \quad L_t=3 \\
K&=&\frac{21}{2}L_t^2-\frac{121}{2}L_t+87, \quad L_t\geq4
\nonumber
\end{eqnarray}
but is anyway limited for practical $L_t$, e.g. $K=13$ with
$L_t=4$.

\item[-] The computation of the $L_t-1$ diagonal terms $r_{2k-1}$
(\ref{r_expr}), with $k=2, \ldots L_t$ requires $2L_t^2-4L_t+3$
RMs.
\end{description}
Overall, the resulting complexity associated with the computation
of the matrix $\tilde{\textbf{R}}$ can be estimated as
$O(2L_rL_t^2+2L_t^2-4L_t+3+K)$.

\noindent$\bullet$ \textit{Observation dependent terms.}

\noindent It should be noted that the explicit computation of the
orthogonal matrix \textbf{Q} is not required, but rather it is
possible to proceed to a direct computation of the elements of the
vector $\tilde{\textbf{y}_r}=\textbf{Q}^T\textbf{y}_r$. In fact
the scalar products $\textbf{q}_j^T\textbf{y}_r$ ultimately depend
on a linear combination of the scalar products
$V_k=\textbf{h}_k^T\textbf{y}_r$, with $k=1, \ldots 2L_t$, whose
total complexity is $4L_tL_r$ RMs. The resulting additional
complexity due to the linear combinations can be estimated as:
\begin{eqnarray}
\label{L_expr}
W&=&6, \quad L_t=2 \\
W&=&16, \quad L_t=3 \nonumber \\
W&=&14L_t-26, \quad L_t\geq4 \nonumber
\end{eqnarray}
The total complexity can then be estimated as $O(4L_tL_r+W)$. It
should be observed that the complexity of the
observation-dependent terms is quadratic with the size of the
system, as opposed to the cubic dependence of the channel
related terms, but for static or slowly-varying channels the
involved operations must be updated more frequently than those
related to the channel.

The processing complexity derived so far does not take into
account the extra-complexity arising from computing some of the
coefficients of the matrix $\tilde{\textbf{R}}$ and the elements
of $\tilde{\textbf{y}_r}$ $L_t$ times (cfr. Section
\ref{sec_LLR_NTX}). A precise complexity estimation would
dependent upon the specific adopted permutation set, of which
(\ref{perms}) is an example. Here we just point out that even in a
pessimistic scenario where no re-use of the formerly executed
computations were possible the resulting complexity would be given
by $L_t$ times $K$ (\ref{K_expr}), $W$ (\ref{L_expr}), and the
number of multipliers associated with the elements $r_{2k-1}$. The
complexity order of magnitude thus would still remain cubic with
the dimension of the MIMO system. It should be stressed that this
is a consequence of not having the normalizations in the GSO.
Thanks to this variation, ultimately the scalar products between
$2L_r$-element vectors which represent the main contribution to
the preprocessing complexity, involve non-normalized channel
columns as in $\textbf{h}_k^T\textbf{y}_r$ or
$\textbf{h}_k^T\textbf{h}_j$. This means they can be re-used in
computing the GSO for any layer ordering.
\newline
\noindent$\bullet$ \textit{Complexity of the lattice search.}

\noindent The complexity associated with the demapping and bit LLR
calculation has a crucial role for hardware implementations of the
algorithm, as the related operations need to be updated for every
channel observation and are proportional to both $L_t$ and
$S=M^2$, the size of the complex constellation. A high-level
estimation can be carried out recalling that the computation of
the bit LLRs corresponding to the $j$-th symbol (\ref{LLR_X_j})
requires $M^2$ squared norms of $2L_t$-element vectors. Thus, in
first approximation the complexity for the whole transmit sequence
is $O(2L_t^2M^2)$ RMs. This estimate is correct under the
assumption that in (\ref{LORDmetrNtx}) the number of products
mainly derives from $2L_t$ squares. That can be justified as
integer $M$-PAM values $x_k$ are to be spanned; thus products like
$cx_k$ where $c$ is a constant value can be handled as sums like
\begin{equation}
cx_k=cx_0+2kc, \nonumber
\end{equation}
where $x_0=-(M-1)$, $k=0,1,\ldots (M-1)$, provided that
intermediate products terms $cx_0$ are stored.

This complexity estimate could be further reduced if
implementation optimizations already proposed for SD
\cite{BBWZFB:05} are adopted for LORD too. Among others, it has to
be mentioned the possibility of "tree pruning", i.e. during PED
term computations (\ref{LORDmetrNtx}) it is possible to take into
account a threshold derived from former EDs and stop the
computation at any layer if the sum of the already computed PEDs
is higher than such a threshold. Besides it should be noted that
possible simplifications to the vector norm computation (e.g.
through $l^1$  or $l^\infty$ norms) may be applied to LORD as
well. However their impact on LLR accuracy should be carefully
evaluated first.

\section{Simulation results} \label{sec_performance}

\begin{figure}
\centering
\includegraphics[width=6.5in]{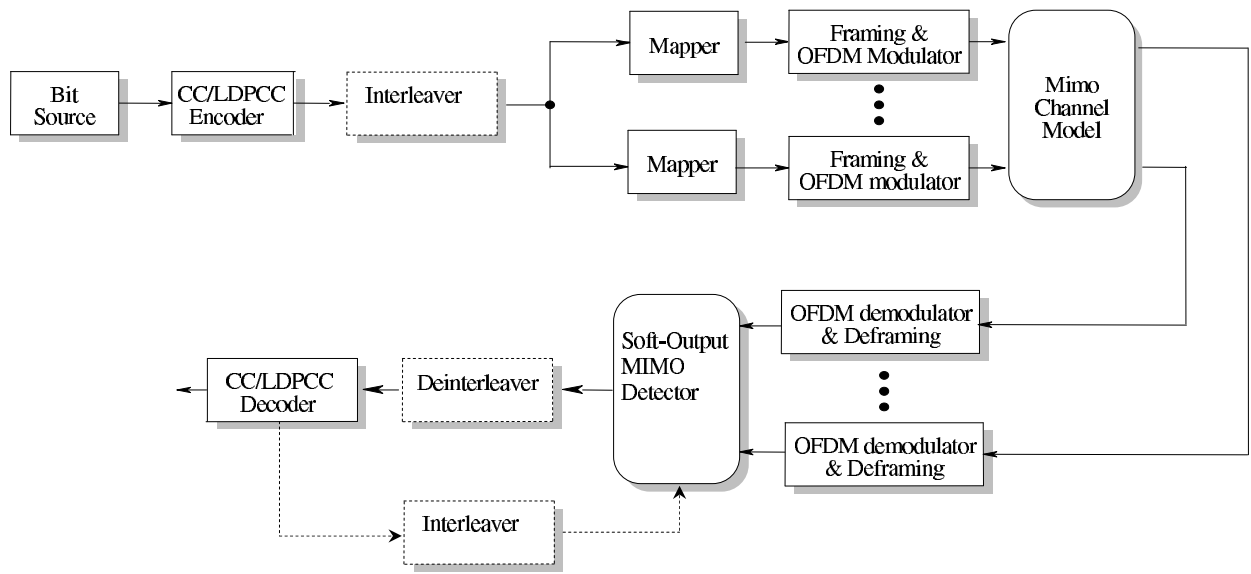}
\caption{MIMO-OFDM BICM block diagram.} \label{fig:mimo_ofdm_bicm}
\end{figure}

\begin{figure}
\centering
\includegraphics[width=5.3in]{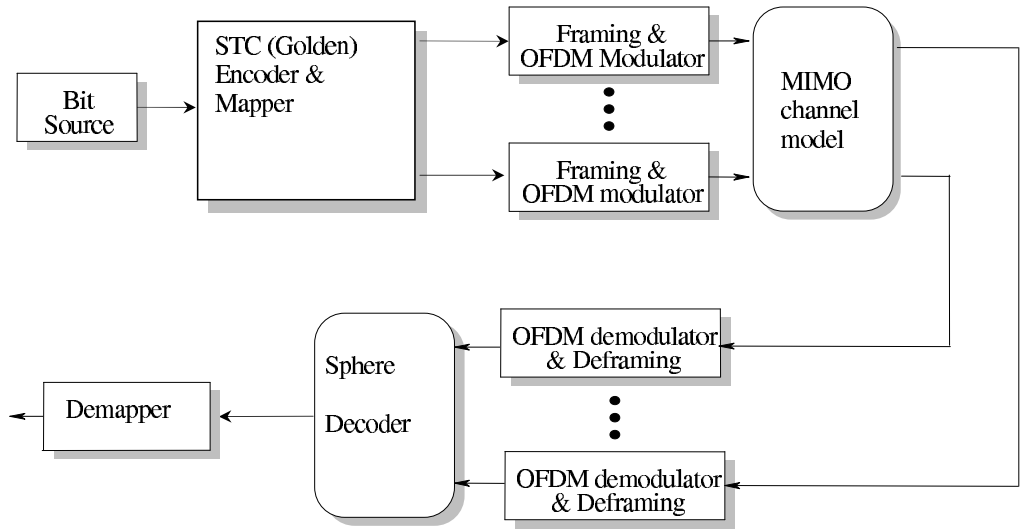}
\caption{MIMO-OFDM STC block diagram.} \label{fig:mimo_ofdm_gc}
\end{figure}

In this section the performance of LORD is reported in two main
MIMO-OFDM configurations of interest: BICM, which is the main
scheme considered by next generation wireless standards (Fig.
\ref{fig:mimo_ofdm_bicm}); STC mapping without concatenated ECC
(Fig. \ref{fig:mimo_ofdm_gc}).

Several detection algorithms have been simulated in MIMO-OFDM BICM
scheme. In a subset of cases, also the performance of
exhaustive-search ML detection was verified, including $L_t=3$ and
16QAM corresponding to 4096 operations per complex symbol. It
should be noted that by ML "exhaustive search over the
constellation symbols" is meant throughout this section.

The block diagram of the system is depicted in Fig.
\ref{fig:mimo_ofdm_bicm}. The system specifications are described
in \cite{80211n:1stWWiSEspec} and represent one of the proposals
for the ongoing standardization activity of IEEE 802.11n next
generation WLANs. In particular, the OFDM parameters are: 54 data
tones out of a total of 64 tones; 20 MHz bandwidth; 3.2 $\mu s$
IFFT/FFT period and 0.8 $\mu s$ guard interval duration. The basic
ECC scheme we considered is a convolutional code (CC) cascaded
with a bit interleaver. The CC decoder is either a soft-input
Viterbi algorithm (VA), or optionally a soft-in soft-out VA (SOVA,
\cite{HH:89}) for use in turbo iterative combined decoder and
detection schemes. CC performance is also compared with an
advanced ECC option, i.e. a low density parity check code (LDPCC);
no interleaver was used in this case. The considered LDPCC
matrices are specified in \cite{80211n:1stWWiSEspec} (1944-bit
coded block size, code rates 1/2, 2/3, 3/4, 5/6); a theoretical
description of their structure can be found in \cite{VCWW:04}.
LDPCC simulations refer to 12 iterations of a log-domain version
of the sum-product algorithm.

In order to verify the performance in different channel conditions
of practical interest, two different frequency selective channel
models \cite{80211TGn:chan} were considered: channel B,
characterized by a 9-tap tapped delay line profile with 15 ns root
mean square (rms) delay spread; channel D, 18-tap and 50 ns rms
delay spread. Channel B is a useful benchmark for scenarios with
limited frequency selectivity, like home residential environment,
while Channel D has a significant frequency diversity as typical
of indoor office.

The performance has been simulated in terms of packet error rate
(PER) versus SNR, for a 1000-byte WLAN packet length; in the
following, SNR gain will be related to $10^{-2}$ PER unless
otherwise stated. The MIMO detector in Fig.
\ref{fig:mimo_ofdm_bicm} operates at subcarrier level, assuming
known channel state information (CSI) and ideal synchronization.
The following soft-output algorithms have been considered: LORD
with max-log bit LLR computation; MMSE with max-log bit LLR
computed taking into account the Gaussian approximation
\cite{ZWDV:03}; iterative MMSE and soft IC (SIC) as in
\cite{SH:02},\cite{ZWDV:03}, with SOVA (optional feedback path in
Fig. \ref{fig:mimo_ofdm_bicm}); exhaustive-search ML with optimal
bit LLR computed through the Jacobian logarithm (or ``$\max^*$"
function) \cite{RVH:95}. MMSE-SIC plots refer to four stages of
MMSE processing (i.e. three iterations), as no appreciable
performance improvement can be observed for more loops.

\begin{figure}
  \centering{
    \subfigure[16QAM code rate 1/2]{\includegraphics[height=3.2in]{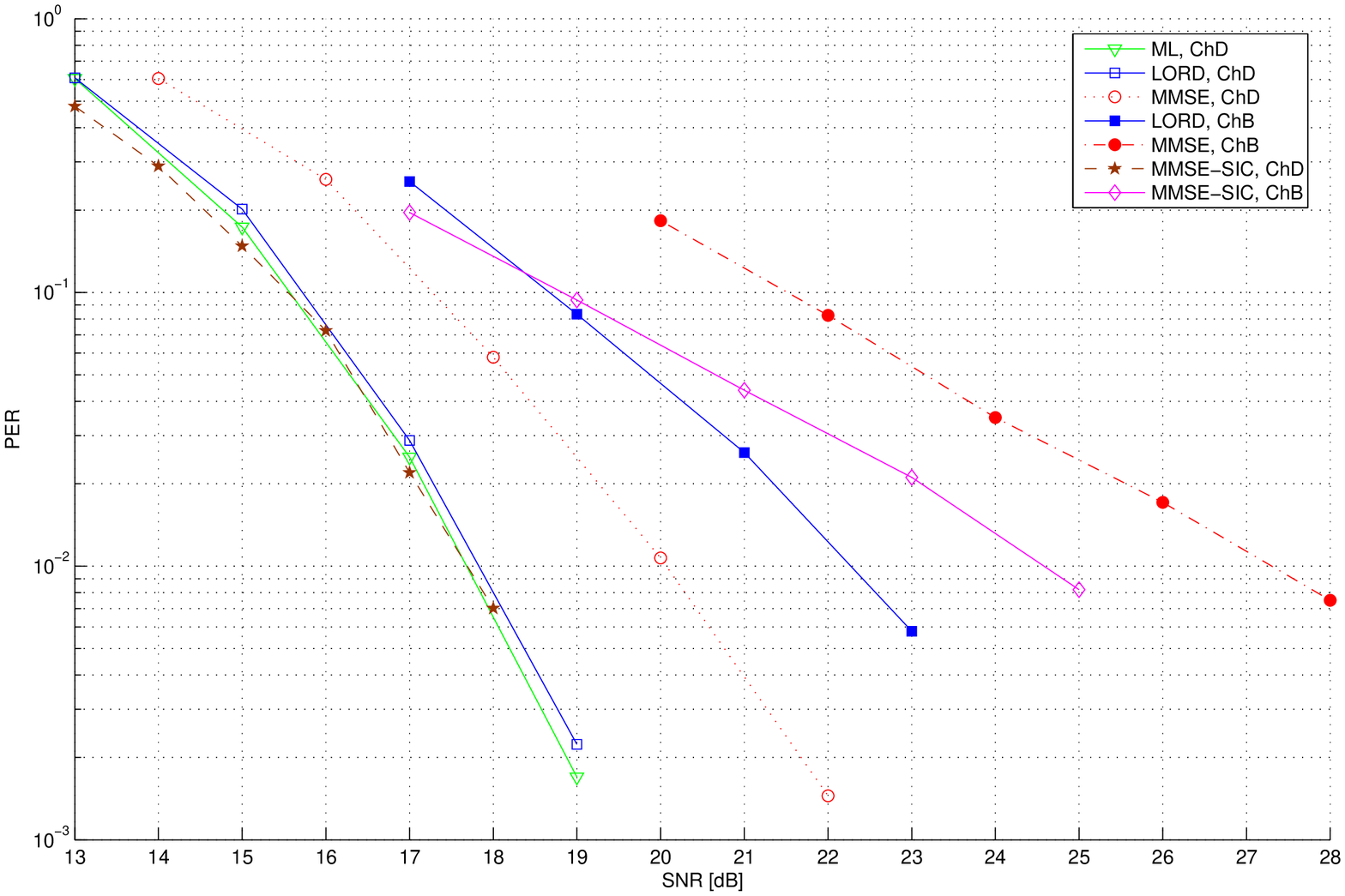}
    \label{2x2a}}
    \hfil
    \subfigure[64QAM code rate 5/6]{\includegraphics[height=3.2in]{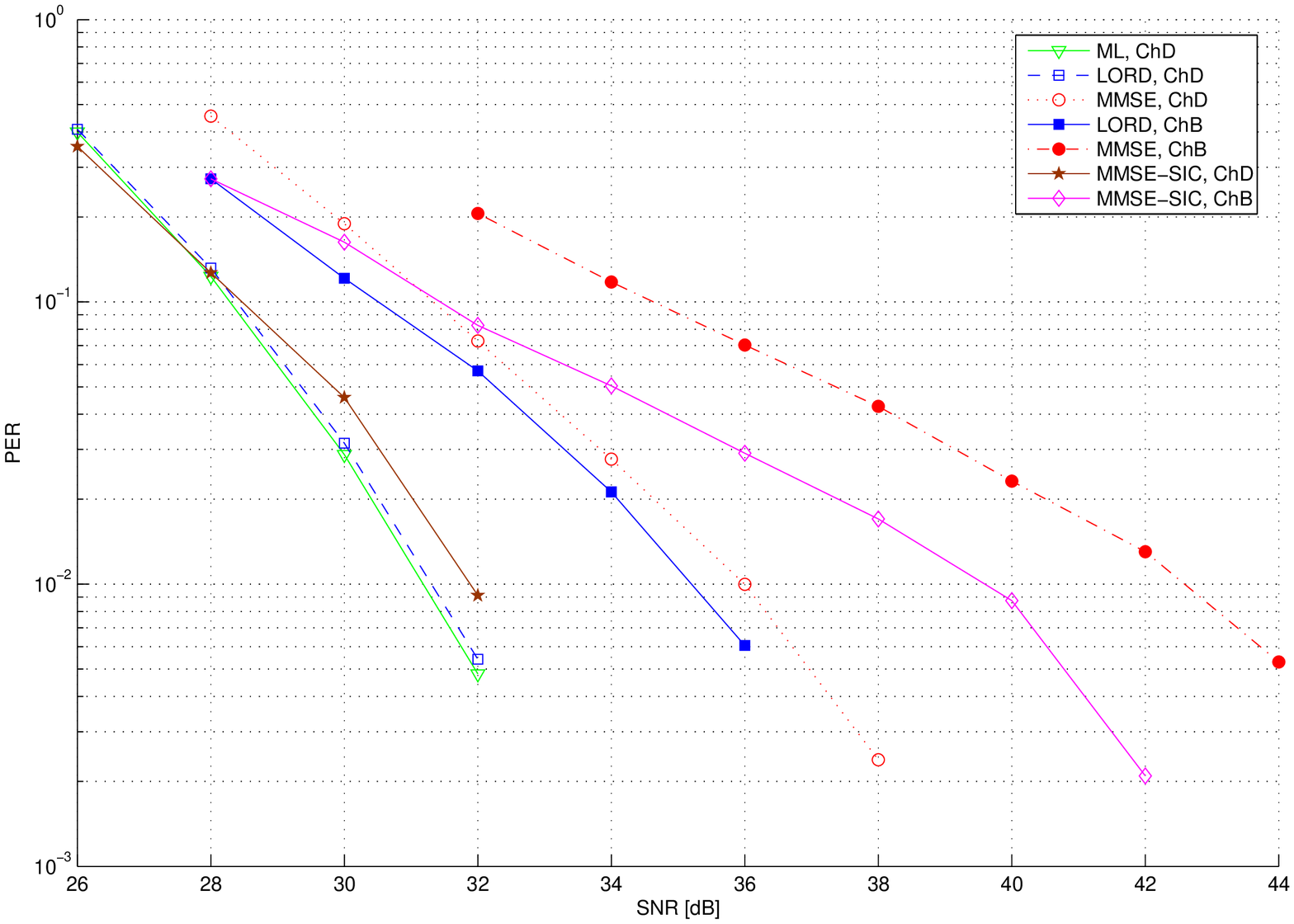}
    \label{2x2b}}}
  \caption{Performance comparison of detection algorithms. $L_t=2$, $L_r=2$ antennas, BICM MIMO-OFDM, convolutional code, channel B and D.}
  \label{fig:2x2perf}
\end{figure}

Fig. \ref{fig:2x2perf} reports the performance of LORD versus MMSE
for channel models B and D, CC coded system, and $L_t=2$, $L_r=2$
(in short, 2x2) MIMO system. Fig. \ref{2x2a} and \ref{2x2b} refer
to 16QAM modulation code rate (CR) 1/2 and 64QAM CR 5/6
respectively. A significant SNR gain over MMSE is visible in all
cases, from a minimum of 2.2 dB for 16QAM modulation CR 1/2 and
channel D, to a maximum of 7.4 dB for 64QAM CR 5/6 and channel B;
also, comparisons with ML confirm the optimality of LORD with
$L_t=2$. The small performance degradation of LORD has to be
attributed to the log-MAP LLR computation used for ML as opposed
to the max-log used for LORD. Interestingly, LORD and MMSE-SIC
show comparable performance in case of channel D, 16QAM CR 1/2
while even a single stage of LORD gains 0.6 dB of SNR over
MMSE-SIC in case of 64QAM CR 5/6. However LORD gains more than 2
dB compared to MMSE-SIC in case of channel B, 16 QAM CR 1/2 and
the advantage increases to about 4.5 dB with 64QAM CR 5/6. These
performance results offer several lines of interpretation. In
terms of CR, the gain of LORD versus a linear suboptimal detector
like MMSE increases for higher CRs. The advantage of LORD is also
significantly higher when less frequency selectivity is made
available by the system, as clear comparing performance obtained
with channel models B and D. In particular, if limited frequency
diversity exists as with channel B, MMSE-SIC does not show an
appreciable BER curve slope improvement compared to a single stage
of MMSE, which is the reason why LORD shows a higher gain in this
condition.

\begin{figure}
\centering
\includegraphics[width=5.4in]{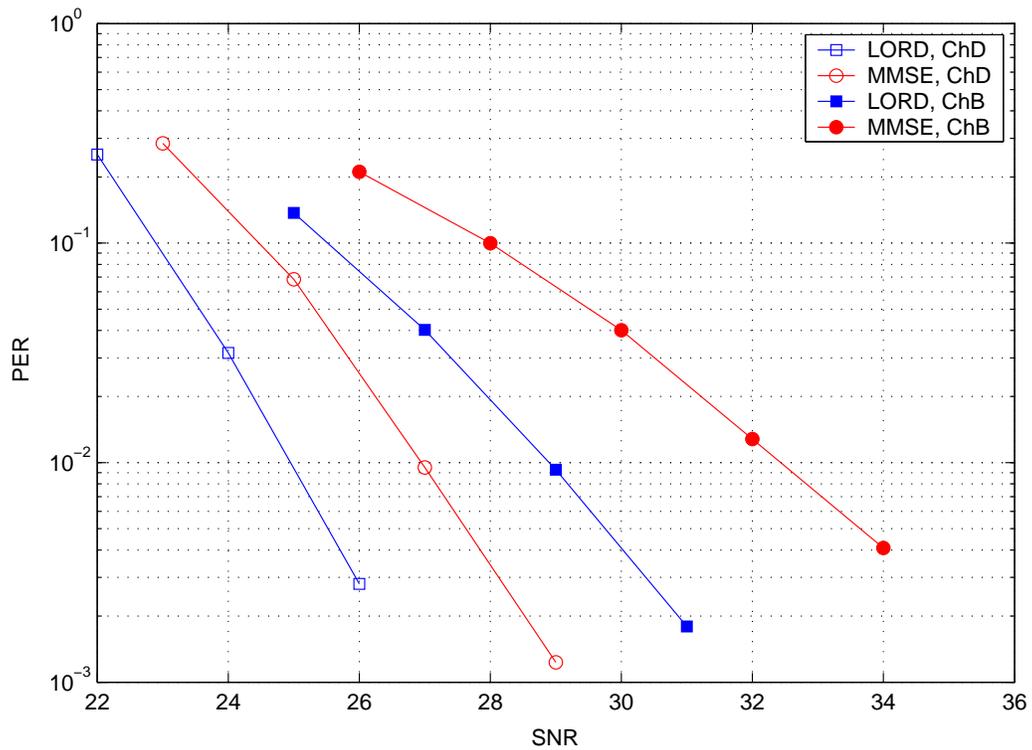}
\caption{Performance comparison of detection algorithms. 64QAM CR
5/6, $L_t=2$, $L_r=3$, BICM MIMO-OFDM, convolutional code,
channels B, D.} \label{fig:2x3perf}
\end{figure}

The performance of CC coded 64QAM, CR 5/6 is shown in Fig.
\ref{fig:2x3perf} in case of a 2x3 MIMO system. It can be noted
that the general trend visible in Fig. \ref{2x2b} still holds also
if additional spatial diversity is made available by the system,
even though the relative gain of LORD versus MMSE decreases;
nevertheless, a SNR gain higher than 3 dB is observable for 64QAM
CR 5/6 and channel model B.

\begin{figure}
  \centering{
    \subfigure[16QAM code rate 1/2]{\includegraphics[height=3.2in]{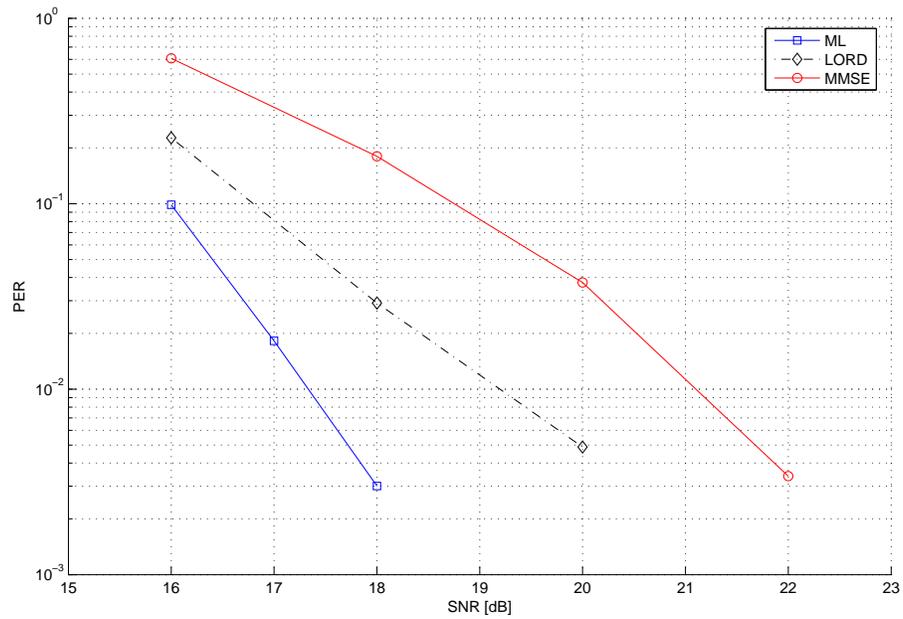}
    \label{3x3a}}
    \hfil
    \subfigure[16QAM code rate 3/4]{\includegraphics[height=3.2in]{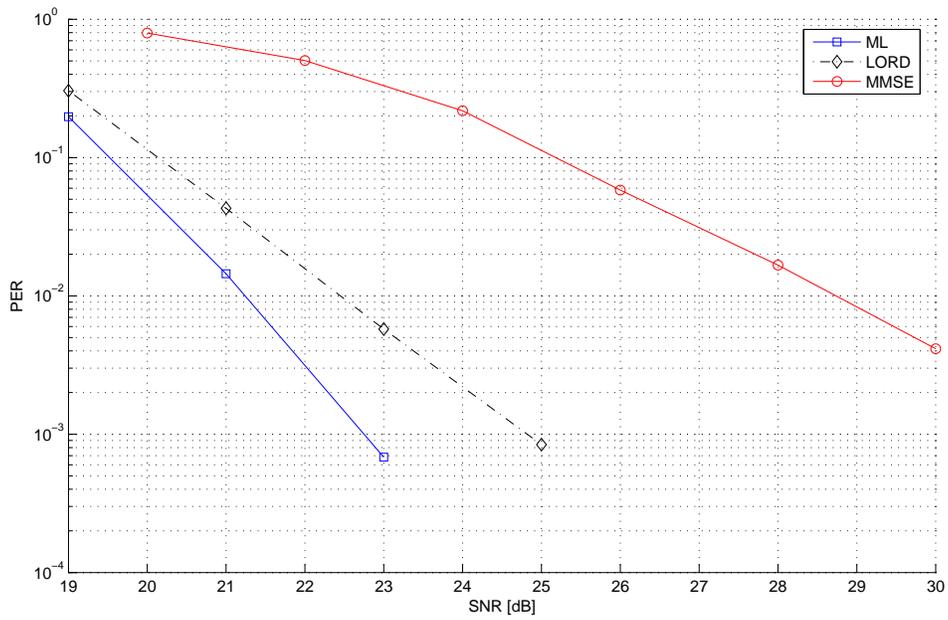}
    \label{3x3b}}}
  \caption{Performance comparison of detection algorithms. $L_t=3$, $L_r=3$, BICM MIMO-OFDM, convolutional code, channel model D.}
  \label{fig:3x3perf}
\end{figure}

\begin{figure}
\centering
\includegraphics[width=5.4in]{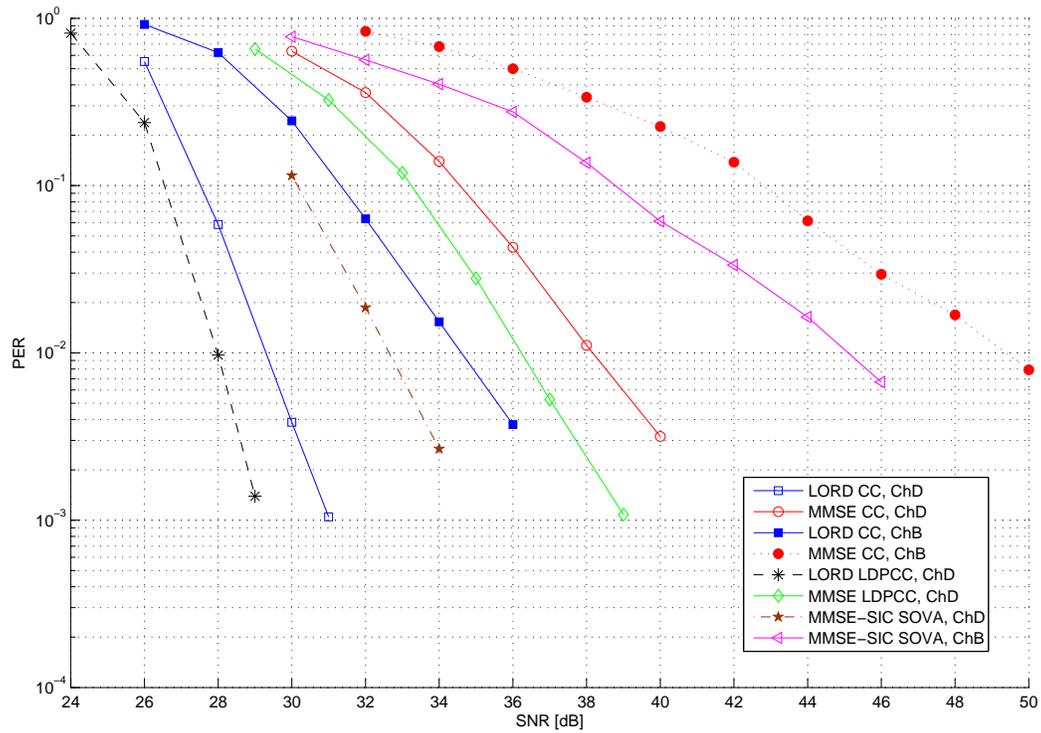}
\caption{Performance comparison. $L_t=4$, $L_r=4$, 64QAM CR 5/6,
BICM MIMO-OFDM, CC and LDPCC, channel B and D.}
\label{fig:4x4perf}
\end{figure}

Fig. \ref{fig:3x3perf}-\ref{fig:4x4perf} show that the advantage
of LORD versus MMSE increases for MIMO systems with a
higher number of transmit antennas (at least up to $L_t=4$). Fig.
\ref{fig:3x3perf} reports the performance of 3x3 16QAM modulation,
CR 1/2 and 3/4 respectively, for ML, LORD and MMSE detectors.
Results confirm that LORD is suboptimal if more than two transmit
antennas are used, but the gap over ML is contained within 2 dB
for CR 1/2, and is only 1.2 dB for CR 3/4; in this last case the
gain over MMSE is about 7.2 dB.

Fig. \ref{fig:4x4perf} summarizes the performance of LORD and MMSE
in case of 4x4 MIMO system, 64QAM modulation, CC coded system with
CR 5/6, channel models B and D; also, two plots with LDPCC and
channel D are provided for comparison. The gain of LORD over MMSE
is 8.9 dB and 14.8 dB with channel model D and B respectively.
Also, LORD shows $>3$ dB of SNR gain over MMSE-SIC with channel D,
while this gain increases to $>9$ dB if channel B is modelled.
Then, LORD can avoid using iterative MMSE detectors, characterized
by latency and complexity disadvantages. LORD iterative schemes
are an envisioned topic for future work.

We note that even though SNR levels much higher than 30 dB are
probably difficult to achieve in practical 802.11n systems, these
results demonstrate the importance of a ML-approaching MIMO
equalizer like LORD in order to implement the highest data rate
transmission schemes currently under definition by 802.11n
standardization committee (the system of Fig. \ref{fig:4x4perf}
corresponds to a data rate of 270 Mbits/s
\cite{80211n:1stWWiSEspec}). This is particularly evident for
channel model B, where MMSE has a dramatical performance
degradation. It should also be observed that an advanced ECC as
LDPCC is able to provide a gain over CC in the order of 2 dB if
used with MMSE, and of 1.2 dB with LORD. The preliminary
conclusion that can be drawn is that an advanced ECC in
combination with a linear detector is not enough to recover the
performance degradation of the system when limited frequency
diversity is present. In this case, also iterative detection
techniques do not prove to be effective if a detector unable to
take advantage of receive diversity, as MMSE, is used as a first
stage.

\begin{figure}
\centering
\includegraphics[width=5.5in]{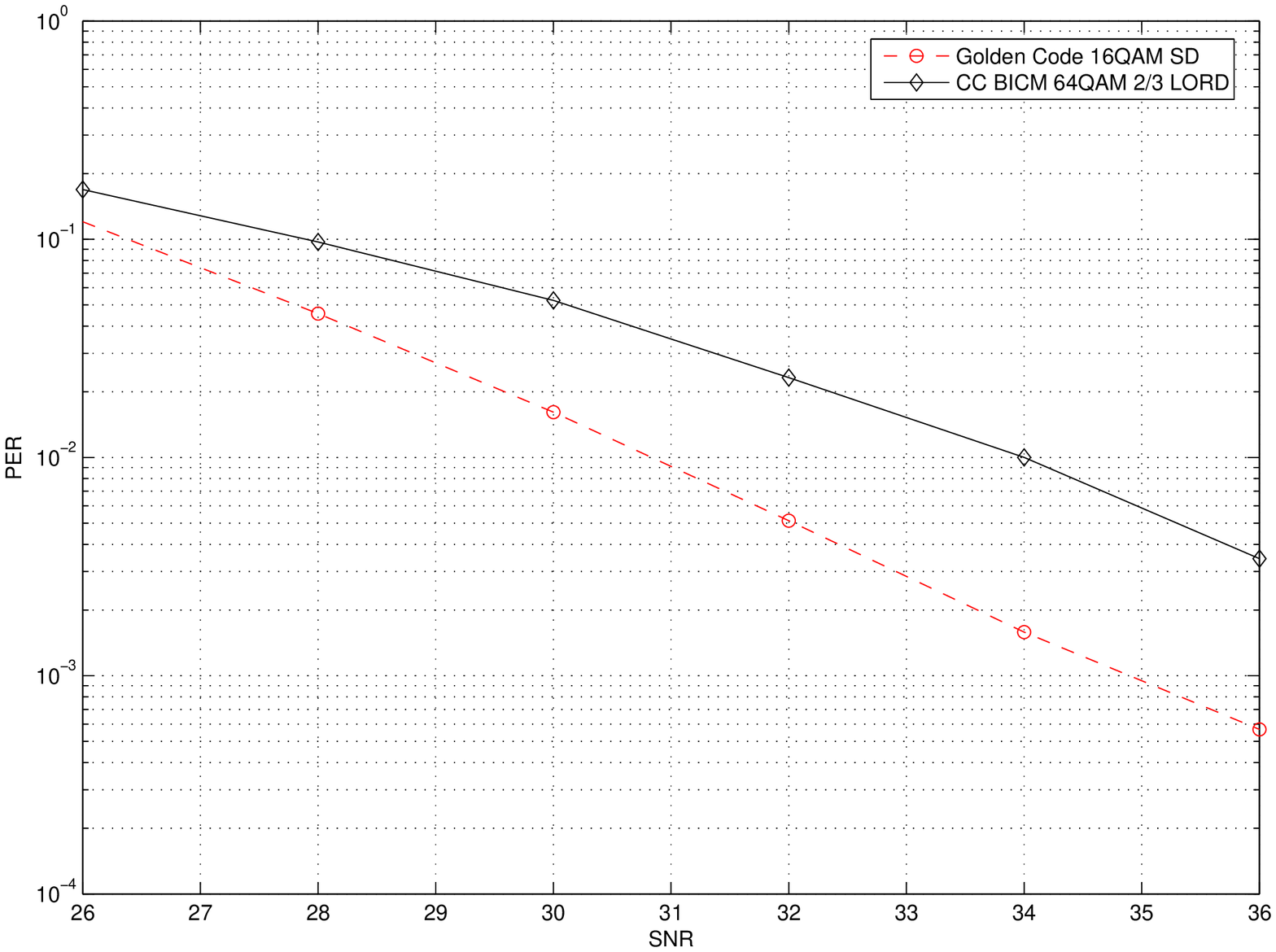}
\caption{Performance comparison of MIMO-OFDM GC, Sphere Decoded,
and CC BICM, 8 bpcu, channel model A.} \label{fig:GC_perf_chA}
\end{figure}

\begin{figure}
\centering
\includegraphics[width=5.5in]{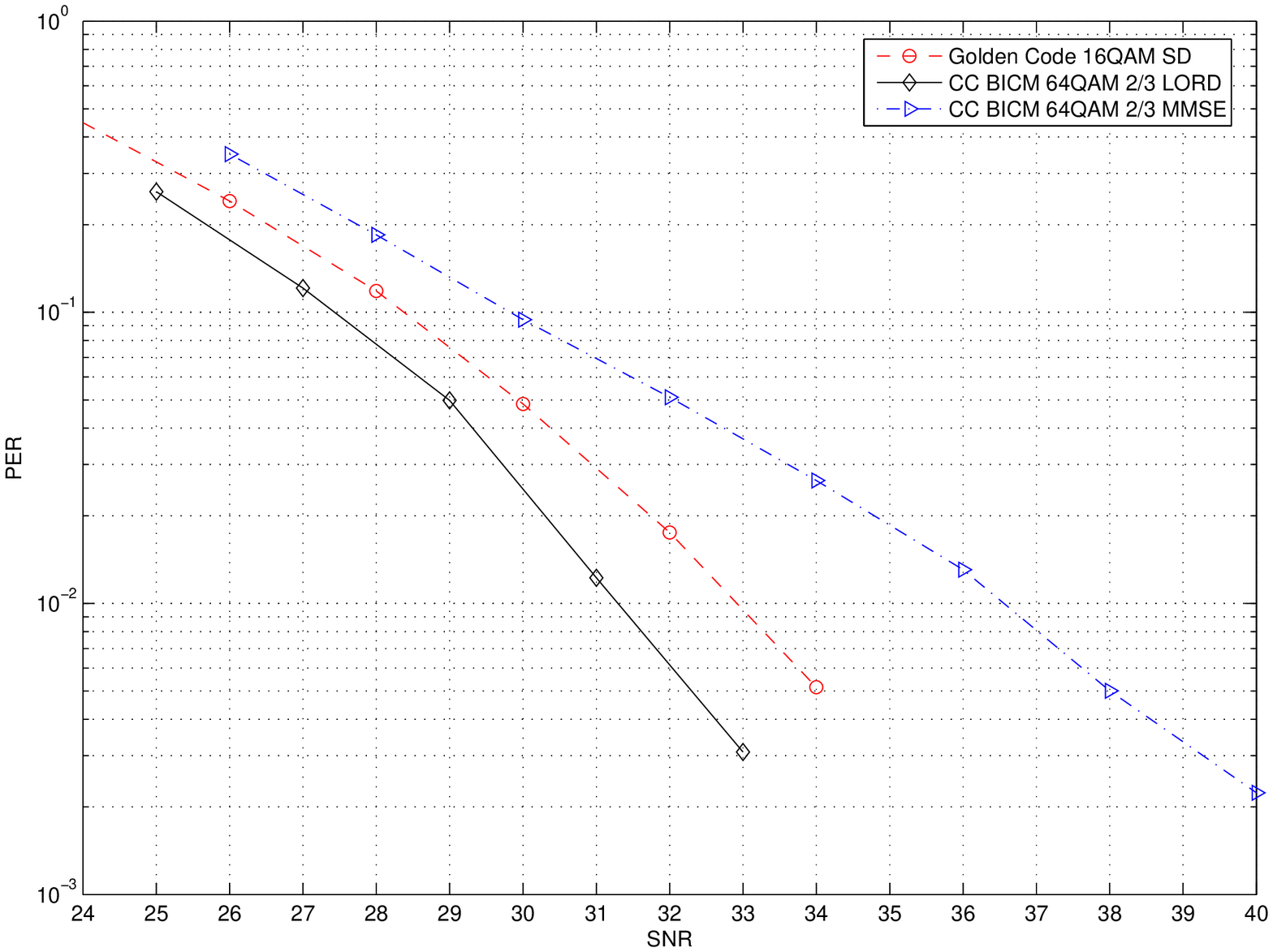}
\caption{Performance comparison of MIMO-OFDM GC, Sphere Decoded,
and CC BICM, 8 bpcu, channel model B.} \label{fig:GC_perf_chB}
\end{figure}

\begin{figure}
\centering
\includegraphics[width=5.5in]{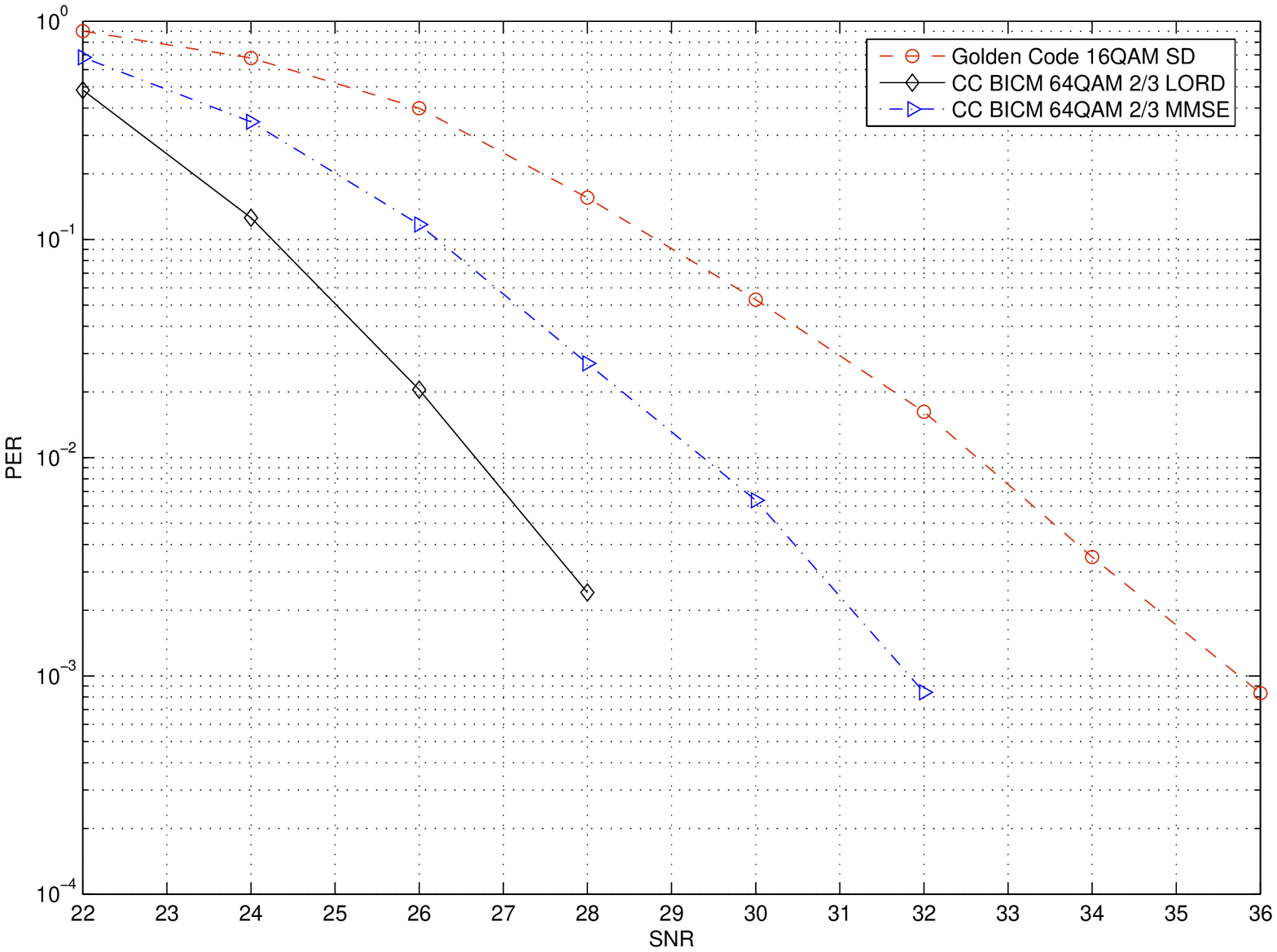}
\caption{Performance comparison of MIMO-OFDM GC, Sphere Decoded,
and CC BICM, 8 bpcu, channel model D.} \label{fig:GC_perf_chD}
\end{figure}

Another case of interest, as previously mentioned, is represented
by the algebraic STCs (ASTCs) \cite{DTB:02},\cite{BRV:05} in
MIMO-OFDM schemes. The interest in this class of codes is
motivated by their ability to yield full data rate (i.e. they
transmit 2 symbols per channel use as $L_t=2$) and a maximal
diversity order $2L_r$ at the same time. Particularly the Golden
Codes (GCs) \cite{BRV:05} outperform all the other classes of STCs
proposed so far to the authors' knowledge. However, ASTCs would
require ML detection in order to provide full diversity order. In
our simulations GCs were decoded using hard-output SD, according
to the MIMO-OFDM block diagram shown in Fig.
\ref{fig:mimo_ofdm_gc}. The OFDM specifications were the same of
the BICM system \ref{fig:mimo_ofdm_bicm}. MIMO-OFDM BICM
LORD-detected systems outperform uncoded GCs for the same bits per
channel use (bpcu), under channel conditions characterized by some
degree of frequency selectivity like channel models B and D
\cite{80211TGn:chan}, as evidenced in Fig. \ref{fig:GC_perf_chB}
and Fig. \ref{fig:GC_perf_chD} respectively for 2x2 and 8 bpcu. It
should be noted that these results were obtained with 64-state CC
Viterbi decoded, i.e. no powerful ECC was necessary. Only in
i.i.d. flat fading channel (model A) the space-time coded system
shows better performance than the BICM system (Fig.
\ref{fig:GC_perf_chA}). Two main considerations can be drawn from
these results. On one hand, they confirm the importance of a
low-complexity soft-output near-ML detector like LORD in order to
fully exploit the space-frequency diversity embedded in layered
BICM systems as specified by practical applications like 802.11n.
On the other hand, the plots also show that for short block length
codes like the GCs, a hard-output ML decoder like SD is not
sufficient to make them attractive for next generation wireless
systems; this would still be true even if optional front-end to
accelerate the decoder convergence were used, as proposed in
\cite{MGDC:05}. We then infer that low-complexity soft-output
near-ML detectors and a properly designed BICM scheme would be
needed also for GCs in MIMO-OFDM schemes. No such detectors have
been proposed so far for full diversity full data rate ASTCs; in
\cite{GC:04} BICM TAST was decoded through a low complexity
message passing iterative decoder. The adaptation of LORD to
optimally decode such codes is considered a topic for future
research. Advanced ECCs do not seem to be essential if enough
frequency selectivity is present in the system.

\section{Conclusions} \label{sec_conclusion}
In this paper a novel MIMO detection algorithm was proposed. LORD
belongs to the class of lattice detector algorithms, though it
uses a novel lattice formulation. A low-complexity channel
preprocessing algorithm was described, alternative to the standard
QR decomposition. The symbol sequence estimation is then performed
through a parallelizable, reduced size and deterministic lattice
search, also suitable for generation of reliable max-log bit LLRs.
LORD was shown to be (max-log) optimal for two transmit antennas
and near-optimal for three and four transmit sources in MIMO-OFDM
BICM configuration, achieving higher SNR gain than linear and
iterative nonlinear detectors, thanks to its very good
exploitation of receive diversity. Also, the performance
comparison with full diversity order two-transmit antenna uncoded
STCs like the GCs showed that LORD detected layered BICM systems
perform better even in presence of simple error correction codes
like a Viterbi decoded convolutional code, provided that some
degree of frequency selectivity characterizes the channel.

\appendix[Recursive formulation of the preprocessing]
\renewcommand{\theequation}{A.\arabic{equation}}

The formulation for the frontend processing that was presented in
Section \ref{sec_preproc_NTX1} can be given in an alternative
equivalent recursive formulation. Recall the observations of
interest are
\begin{equation}
\tilde{\textbf{y}}_r={\bf Q}^T
\textbf{y}_r=\sqrt{\frac{E_{s}}{L_t}} \, {\bf \tilde{R}}
\textbf{x}_r + {\bf Q}^T \textbf{n}_r=\sqrt{\frac{E_{s}}{L_t}} \,
\tilde{{\bf R}} \textbf{x}_r + \tilde{\textbf{n}}_r
\label{sys_app}
\end{equation}
The orthogonal matrix can be obtained by defining the following
quantities:
\begin{equation}
\textbf{e}_I(0,j)=\textbf{h}_{2j-1} \qquad
\textbf{e}_Q(0,j)=\textbf{h}_{2j} \qquad
\sigma(0,j)=\left|\textbf{h}_{2j-1}\right|^2 \qquad 1 \le j \le
L_t
\end{equation}
with the following three order recursions ($1 \le i<j \le L_t$)
\begin{eqnarray}
\textbf{e}_I(i,j)&=&\sigma(i-1,i)\textbf{e}_I(i-1,j)
-r_I(i,j)\textbf{e}_I(i-1,i)-r_Q(i,j)\textbf{e}_Q(i-1,i) \label{eq:recuri}\\
\textbf{e}_Q(i,j)&=&\sigma(i-1,i)
\textbf{e}_Q(i-1,j)+r_Q(i,j)\textbf{e}_I(i-1,i)-r_I(i,j)\textbf{e}_Q(i-1,i) \label{eq:recurq} \\
\sigma(i,j)&=&\sigma(i-1,i)\sigma(i-1,j)-\left(r_I(i,j)\right)^2
-\left(r_Q(i,j)\right)^2
\end{eqnarray}
where
\begin{equation}
r_I(i,j)=\textbf{e}_I(i-1,i)^T\textbf{e}_I(0,j)  \qquad
r_Q(i,j)=\textbf{e}_Q(i-1,i)^T\textbf{e}_I(0,j).
\end{equation}
With these definitions in place the columns of the orthogonal
matrix are defined with vectors $\textbf{b}_i$
\begin{equation}
\textbf{b}_{2i-1}=\textbf{e}_I(i-1,i) \qquad
\textbf{b}_{2i}=\textbf{e}_Q(i-1,i) \qquad 1 \le i \le L_t.
\end{equation}
Computing the $i^{\mbox{\scriptsize th}}$ pair of orthogonal
vectors would require $i-1$ recursions of (\ref{eq:recuri}) and
(\ref{eq:recurq}), each one involving $2L_r$ terms. However, from
(\ref{sys_app}) and as observed in Section \ref{sec_complexity},
the matrix ${\bf Q}$ does not actually have to be computed to
accomplish detection. The vectorial recursions specified above are
important in computing terms that only appear in scalar products
which, in their turn, can be expressed as linear combinations of
the initialization scalar vectors
\begin{equation}
r_{I0}(i,j)=\textbf{e}_I(0,i)^T\textbf{e}_I(0,j)  \qquad
r_{Q0}(i,j)=\textbf{e}_Q(0,i)^T\textbf{e}_I(0,j) \qquad 1 \le i <
j \le L_t.
\end{equation}
The upper triangular matrix that results from the decomposition is
also simply specified. The diagonal elements of ${\bf \tilde{R}}$
have the form:
\begin{equation}
\tilde{R}_{2i-1,2i-1}=\sigma(i-1,i)=\tilde{R}_{2i,2i}.
\end{equation}
The upper triangular elements are:
\begin{equation}
\tilde{R}_{2i-1,2j-1}=r_I(i,j) \qquad \tilde{R}_{2i-1,2j}=r_Q(i,j)
\qquad 1 \le i < j \le L_t
\end{equation}
and
\begin{equation}
\tilde{R}_{2i,2j-1}=-r_Q(i,j) \qquad \tilde{R}_{2i,2j}=r_I(i,j)
\qquad 1 \le i < j  \le L_t.
\end{equation}
The noise, $\tilde{\textbf{n}_r}$, remains white and has a
component-wise variance given as
\begin{equation}
R_{\tilde{n}_{2i-1,2i-1}}=R_{\tilde{n}_{2i,2i}}=\frac{N_0}{2}
\prod_{j=1}^{i} \sigma(j-1,j).
\end{equation}
These recursions give the components of the upper triangular model
that are needed for the detection algorithm.

The post-processed observations are also specified with a
recursion. These recursions are given as
\begin{eqnarray}
y_I(i,j)&=&\sigma(i-1,i)y_I(i-1,j)-r_I(i,j)y_I(i-1,i)-r_Q(i,j)y_Q(i-1,i)\\
y_Q(i,j)&=&\sigma(i-1,i)y_Q(i-1,j)+r_Q(i,j)y_I(i-1,i)-r_I(i,j)y_Q(i-1,i)
\end{eqnarray}
with the following initial conditions
\begin{equation}
y_I(0,j)=\textbf{e}_I(0,j)^T \textbf{Y}_r \qquad
y_Q(0,j)=\textbf{e}_Q(0,j)^T \textbf{Y}_r.
\end{equation}
The final outputs for the detection will be
\begin{equation}
\left[\textbf{y}_r\right]_{2i-1}=y_I(i-1,i) \qquad
\left[\textbf{y}_r\right]_{2i}=y_Q(i-1,i) \qquad 1 \le i \le L_t.
\end{equation}
In examining the recursions needed for the upper triangular model
parameters and the observations it is apparent that the orthogonal
vector pairs, $\textbf{q}_{2i-1}$ and $\textbf{q}_{2i}$, only have
to be computed up to the $i\le L_t-1$ level.

A couple examples will help illustrate the proposed recursions.
First consider again the $L_t=2$ case and here the preprocessing
algorithm has the following pseudo code
\begin{enumerate}
\item  Layer 1
\begin{enumerate}
\item Compute $\sigma(0,1)$ -- Complexity $O(2L_r)$, \item
Compute $y_I(0,1)$ -- Complexity $O(2L_r)$, \item  Compute
$y_Q(0,1)$ -- Complexity $O(2L_r)$,
\end{enumerate}
\item  Layer 2
\begin{enumerate}
\item  Compute $\sigma(0,2)$ -- Complexity $O(2L_r)$, \item
Compute $y_I(0,2)$ -- Complexity $O(2L_r)$, \item  Compute
$y_Q(0,2)$ -- Complexity $O(2L_r)$, \item  Compute $r_I(1,2)$ --
Complexity $O(2L_r)$, \item  Compute $r_Q(1,2)$ -- Complexity
$O(2L_r)$, \item  Compute $\sigma(1,2)$ -- Complexity $O(3)$,
\item  Compute $y_I(1,2)$ -- Complexity $O(3)$, \item  Compute
$y_Q(1,2)$ -- Complexity $O(3)$.
\end{enumerate}
\end{enumerate}
Again as noted above the algorithm for $L_t=2$ has complexity
$O(16L_r+9)$. For the $L_t=4$ case the preprocessing algorithm has
the following pseudo code
\begin{enumerate}
\item  Layer 1 - Same as $L_t=2$ case, \item  Layer 2 - Same as
$L_t=2$ case, \item  Layer 3
\begin{enumerate}
\item  Initialization $\sigma(0,3), y_I(0,3), y_Q(0,3)$ --
Complexity $O(6L_r)$, \item  First recursion $r_I(1,3), r_Q(1,3),
\sigma(1,3), y_I(1,3), y_Q(1,3)$ -- Complexity $O(4L_r+9)$, \item
Second recursion $r_I(2,3), r_Q(2,3), \sigma(2,3), y_I(2,3),
y_Q(2,3)$ -- Complexity $O(4L_r+15)$,
\end{enumerate}
\item  Layer 4
\begin{enumerate}
\item  Initialization $\sigma(0,4), y_I(0,4), y_Q(0,4)$ --
Complexity $O(6L_r)$, \item  First recursion $r_I(1,4), r_Q(1,4),
\sigma(1,4), y_I(1,4), y_Q(1,4)$ -- Complexity $O(4L_r+9)$, \item
Second recursion $r_I(2,4), r_Q(2,4), \sigma(2,4), y_I(2,4),
y_Q(2,4)$ -- Complexity $O(4L_r+15)$, \item Third recursion
$r_I(3,4), r_Q(3,4), \sigma(3,4), y_I(3,4), y_Q(3,4)$ --
Complexity $O(4L_r+15)$,
\end{enumerate}
\end{enumerate}
The overall complexity for $L_t=4$ case is
$O\left(48L_r+72\right)$. In general the complexity of the
preprocessing algorithm is $O\left(2L_rL_t^2+4L_tL_r\right)$, the
same order of magnitude reported in Section \ref{sec_complexity}.

\section*{Acknowledgment}
The authors wish to thank E. Gallizio, D. Gatti, M. Odoni, A.
Poloni, F. Spalla, A. Tomasoni, and S. Valle, for their
fundamental role in the 802.11n system development.

\bibliographystyle{IEEEtran}
\bibliography{max}
\newpage

\end{document}